\documentclass[a4paper, amsfonts, amssymb, amsmath, reprint, showkeys, nofootinbib, twoside]{revtex4-1}
\usepackage[english]{babel}
\usepackage[utf8]{inputenc}
\usepackage{graphicx}% Include figure files
\usepackage{dcolumn}% Align table columns on decimal point
\usepackage{bm}% bold math
\usepackage{textcomp}
\usepackage{color}
\usepackage{amsmath}
\usepackage[ISO=false]{diffcoeff}
\diffdef{} 
{
    op-symbol = \mathrm{d},
    *-op-left = true,
    op-order-sep = 0 mu,
}

\renewcommand{\b}[1]{\mathbf{#1}}

\newcommand{\var}{\operatorname{var}}
\newcommand{\avg}[1]{\left\langle{#1}\right\rangle}
\newcommand{\qt}{{q_{\theta}}}

\newcommand{\htOi}{\hat\tau^{\mathcal{O},int}}
\newcommand{\htOF}{\hat\tau^{\mathcal{O}}_F }
\newcommand{\htOFi}{\hat\tau^{\mathcal{O},int}_F }
\newcommand{\htB}{\hat{\tau}_B}
\newcommand{\htBi}{\hat{\tau}_B^{int}}

\newcommand{\tL}{\tau_{L}}
\newcommand{\tLi}{\tau_{L}^{int}}
\newcommand{\tLOi}{\tau_{L}^{\mathcal{O},int}}
\newcommand{\tOL}{\tau_{L}^{\mathcal{O}}}
\newcommand{\tOi}{\tau^{O,int}}

\newcommand{\tF}{\hat\tau^{H,int}_F }
\newcommand{\tB}{\hat\tau^{int}_B }
\newcommand{\tI}{\hat\tau^{H,int}}

\newcommand{\dkl}{\hat{D}_{KL}}
\newcommand{\dlam}{\hat{D}_{\lambda_1}}
\newcommand{\dess}{\hat{D}_{ESS}}

\begin{document}

\title{Analysis of autocorrelation times in Neural Markov Chain Monte Carlo simulations}

\author{Piotr Białas}
\email{piotr.bialas@uj.edu.pl}
\affiliation{Institute of Applied Computer Science, Jagiellonian University, ul. \L ojasiewicza 11, 30-348 Krak\'ow, Poland}
\author{Piotr Korcyl}
\email{piotr.korcyl@uj.edu.pl}
\author{Tomasz Stebel}
\email{tomasz.stebel@uj.edu.pl}
\affiliation{Institute of Theoretical Physics, Jagiellonian University, ul. \L ojasiewicza 11, 30-348 Krak\'ow, Poland}

\date{\today}

\begin{abstract}

We provide a deepened study of autocorrelations in Neural Markov Chain Monte Carlo (NMCMC) simulations, a version of the traditional Metropolis algorithm which employs neural networks to provide independent proposals. We illustrate our ideas using the two-dimensional Ising model. We discuss several estimates of autocorrelation times in the context of NMCMC, some inspired by analytical results derived for the Metropolized Independent Sampler (MIS). We check their reliability by estimating them on a small system where analytical results can also be obtained. Based on the analytical results for MIS we  propose a new loss function and study its impact on the autocorelation times. Although, this function's performance is a bit inferior to the traditional Kullback-Leibler divergence, it offers two training algorithms which in some situations may be beneficial.
By studying a small, $4 \times 4$, system we gain access to the dynamics of the training process which we visualize using several observables. Furthermore, we quantitatively investigate the impact of imposing global discrete symmetries of the system in the neural network training process on the autocorrelation times. Eventually, we propose a scheme which incorporates partial heat-bath updates which considerably improves the quality of the training. The impact of the above enhancements is discussed for a $16 \times 16$ spin system. The summary of our findings may serve as a guidance to the implementation of Neural Markov Chain Monte Carlo simulations for more complicated models.
\end{abstract}

\keywords{Autoregressive Neural Networks, Monte Carlo simulations, Ising model, Autocorrelations}

\maketitle

\section{Introduction}

The original idea behind Monte Carlo simulations proposed by Stanislaw Ulam and Enrico Fermi and later expressed in a form of an algorithm applied to study a simple classical statistical mechanics problem by Metropolis et al. \cite{metropolis} turned out to be a very powerful approach to tackle a large variety of problems in all computational sciences (see for example \cite{binder}). In case of complicated probability distributions most formulations resort to the construction of an associated Markov chain of consecutive proposals. The statistical uncertainty of any outcome of Monte Carlo simulation depends directly on the number of statistically independent configurations used to estimate it. Hence, the effectiveness of simulation algorithms is measured by autocorrelation time which quantifies how many configurations are produced by the algorithm before a new, statistically independent configuration appears. Increasing autocorrelation times, a phenomenon called critical slowing down, is usually the main factor which limits the statistical precision of outputs. In the context of field theory simulations in elementary particle physics several proposals were advanced in order to alleviate that problem, in particular metadynamics \cite{metadynamics, metadynamics_nature}, instanton updates \cite{Fucito:1984gw} or multiscale thermalization \cite{PhysRevD.92.114516}.

The recent interest in machine learning techniques has also provided new ideas in the domain of Monte Carlo simulations which aim at reducing the autocorrelation times. The ability of artificial neural networks to emulate a very wide class of probability distributions was used in Ref.\cite{2019PhRvL.122h0602W} in the Variational Autoregressive Network (VAN) approach to approximate free energy of statistical systems. Subsequently the idea was extended and used as a mechanism of providing uncorrelated proposals in a Monte Carlo simulation in Ref.\cite{2020PhRvE.101b3304N}. The Authors call  the resulting algorithm Neural Markov Chain Monte Carlo (NMCMC). The idea to use self-learning neural network as a sampler for MCMC was proposed first in context of field theory in Ref.~\cite{PhysRevD.100.034515}, where Normalizing Flows were used (see also Ref.~\cite{2021arXiv210108176A}). In parallel, the application of  restricted Boltzmann machines to MC was developed in Ref.~\cite{PhysRevB.95.035105}.

Since the proposal of the NMCMC algorithm a few studies discussed its effectiveness. In particular, the Authors of Ref. \cite{2020PhRvE.101e3312M} applied a somewhat similar method based on Neural Autoregressive Distribution Estimators (NADE) to two dimensional Edwards-Anderson model of spin glass. NADE requires a set of data obtained using standard Monte Carlo method in training process. In Ref. \cite{2021arXiv210505650W}  a different sampling mechanism for MCMC was proposed where only a portion of the configuration is updated using a machine learning mechanism and symmetries of the system are utilized. In the present study, we concentrate on the classical test bench provided by the two-dimensional Ising model in order to stay in a simple and well understood  physical situation and focus our attention on the algorithmic effectiveness of the approach. We quantify the latter using several possible measures of autocorrelation times and compare them with the original Metropolis algorithm \cite{metropolis} and with the cluster algorithm proposed by Wolff \cite{PhysRevLett.62.361, WOLFF1989379}.  The Ising model offers also exact solution for the partition function of the finite-size system \cite{PhysRev.185.832} from which the free energy can be obtained and which provides measures of how well the neural network approximates the desired probability distribution.

In the following we start in Section \ref{subsection: ising} by briefly defining the model, in Section \ref{subsection: nmcmc} we present the Neural Markov Chain Monte Carlo (NMCMC) algorithm \cite{2020PhRvE.101b3304N} using Variational Autoregressive Networks. In Section \ref{section: liu} we summarize the known, although so far not discussed in the context of NMCMC, analytical results for the autocorrelation time in a simulation using the Metropolized Independent Sampler (MIS) derived by Liu in Ref. \cite{Liu} and discuss them in the present setup. Subsequently, in Section \ref{sect_autocorr_times_sect} we define several possible estimators for the autocorrelation time. In Section \ref{section: loss functions} we collect two loss functions used in the present context in the Literature and we supplement them with a new loss function inspired by the analytic insight discussed in Section \ref{section: liu}. The results of our numerical experiments are divided into two parts: Sections \ref{section: results small} and \ref{section: results large}. In the former part we use a small system size which enables us to estimate all analytical quantities of interest and provide their comparisons with the approximated counterparts defined on a small number of configurations contained in a batch. In the latter part we use a larger system size in order to check the algorithmic effectiveness for more realistic system sizes. Finally, in the remaining Section \ref{sec: reduction} we describe two techniques which improve the quality of the training of the neural network. First, we exploit the symmetries. In Section \ref{section: symmetries} we discuss the effect of enforcing the $Z_2$ and translational symmetries during the training process. Second, in Section  \ref{section: chessboard}, we use the locality of nearest-neighbour interactions of the Hamiltonian of the Ising model to reduce the number of relevant spin that have to be modelled by the neural network. By using so-called chessboard partitioning half of the spins has to be generated by the neural network while the remaining half can be obtained exactly from the local Boltzmann probability distribution providing the heath-bath type of update. We provide a summary of our results and an outlook in Section \ref{section: summary}.

\subsection{Ising model}
\label{subsection: ising}

We consider the two-dimensional ferromagnetic Ising model with $N=L^2$ spins which are located on the $L\!\!\times\!\! L$ square lattice. We will always compute observables on configurations coming from a Markov chain, hence labelled by a "Monte Carlo" time. We denote the $k$-th configuration in the chain by $\mathbf{s}_k$. The spins within a single configuration are labelled by a linear upper index $i$; thus the $i$-th spin of $k$-th configuration is $s_k^i$. Each state of the system is thus a configuration of spins $\mathbf{s}_k = (s_k^1,\ldots s_k^N)$, and $s_k^i=\pm 1$. For a  configuration of spins $\mathbf{s}_k$ the energy of that configuration is given by the Hamiltonian
\begin{equation}
    H(\mathbf{s}_k) = - J \sum_{\langle i,j \rangle} s_k^i \, s_k^j \,,
\label{Ising_hamilt}
\end{equation}
where the  sum runs over all neighbouring pairs of lattice sites.
In the following, we set the coupling constant $J=1$ without loosing generality. We assume  periodic boundary conditions for the lattice.  

The corresponding Boltzmann distribution at inverse temperature $\beta$ is then given by
\begin{align}
    p(\mathbf{s}_k) = \frac{1}{Z} \exp(-\beta H(\mathbf{s}_k)) \,,
    \label{eq_boltz_distr}
\end{align}
with partition function $Z=\sum_{\mathbf{s}} \exp(-\beta H(\mathbf{s}))$, where the sum is performed over all $M=2^N$  states  of the system.

In the infinite volume limit, the model undergoes second-order phase transition at $\beta_c = \frac{1}{2} \ln (1 + \sqrt{2}) \approx 0.441$. The corresponding order parameter is the absolute magnetization per spin
\begin{multline}
    m = \avg{\frac{1}{N} \left| \sum_{i=1}^N \mathbf{s}^i\right|}_{p}=\sum_{k=1}^{M} p(\mathbf{s}_k)
    \left( \frac{1}{N} \left| \sum_{i=1}^N \mathbf{s}^i_k\right| \right) \\
    \approx \frac{1}{N_{batch}}\sum_{k=1}^{N_{batch}} \frac{1}{N} \left| \sum_{i=1}^N \mathbf{s}^i_k\right|,
\end{multline}
where the last expression is an average over $N_{batch}\gg 1$ configurations sampled from distribution $p$. In a finite volume the system exhibits a crossover at $\beta_c(L)$, i.e. the value of $\beta$ at given $L$ where $|\frac{\partial m}{\partial \beta}|$ is maximal. The finite volume system is in an ordered state when $\beta>\beta_c(L)$ and in a  disordered state for $\beta < \beta_c(L)$.

\subsection{Neural Markov Chain Monte Carlo (NMCMC)}
\label{subsection: nmcmc}
In the VAN approach discussed in this paper, we use an artificial neural network to model the probability $p(\b{s})$. In order to do so we must first recast $p(\b{s})$ in the form that will ensure proper normalisation and allow for easy sampling from the learned distribution. To this end we rewrite  $p(\mathbf{s}_k)$  as a product of conditional probabilities of consecutive spins, i.e.
\begin{multline}
\label{eq:conditional_probabilities}
    p(\mathbf{s}_k) = p(s^1={s}_k^1)\prod_{i=2}^N p(s^i = {s}_k^i | s^1 = {s}_k^1, s^2 = {s}_k^2, \dots, \\ s^{i-1} = {s}_k^{i-1} )
\end{multline}
Note that factorization of the total probability of the configuration into $N$ conditional probabilities requires a unique ordering of all spins on the lattice. However, the precise form of the ordering does not matter if it is maintained during the entire simulation. For definiteness, in our implementation we use a lexicographic ordering on the two-dimensional lattice, the $x$ coordinate being faster than the $y$ coordinate.

Now we can use an artificial neural network to model the functional dependence of all conditional probabilities $p(s^i|\cdot)$. In the simplest realization, the neural network will have $N$ input neurons and $N$ output neurons. The value of the output of the $i$-th neuron is interpreted as 
\begin{equation}
    q_{\theta}(s^i = 1 | s^1, s^2 , \dots, s^{i-1} )
\end{equation}
In order to ensure that the conditional probabilities depend only on the values of the preceding spins a particular architecture of artificial neural networks is used, the so-called autoregressive networks. They are build out of a  class of dense or convolutional layers where the neural connections are only allowed to the neurons associated to the preceding spins. The details of such a construction can be found in Ref.\cite{2019PhRvL.122h0602W}.

It is the purpose of the training steps to tune the  weights of the neural network, collectively denoted as $\theta$ in such a way as to approximate $p(\b{s})$ as close as possible. This is accomplished  by looking for the  minimum of the loss function, usually the Kullback–Leibler divergence which quantifies the difference of two probability distributions. In the following we also discuss other possible choices and their interpretation.

We can easily generate configurations from the $q_\theta(\mathbf{s})$
distribution by generating subsequent spins from conditional probabilities $q_\theta(s^i|s^1,\ldots,s^{i-1})$. This requires $N$ sequential calls to $q_\theta(s|\cdot)$ functions i.e. making  $N$ forward passes through the neural network.  Please note 
that this procedure does not rely in any place on any preceding spin configuration, hence it provides a mechanism for generating uncorrelated configurations from the $q_{\theta}$ distribution avoiding the construction of a Markov chain.

In the end we are interested in spin configurations generated with the target, Boltzmann probability distribution. Having a neural network generating states according to the distribution $q_\theta$ we can correct the difference between $p(\mathbf{s})$ and $q_{\theta}(\mathbf{s})$ within an accept/reject step with the accept probability for the state $\mathbf{s}_{k+1}$,
\begin{equation}
    \min \left( 1, \frac{p(\mathbf{s}_{k+1}) q_\theta(\mathbf{s}_{k})}{p(\mathbf{s}_{k}) q_\theta(\mathbf{s}_{k+1}) }   \right) 
    . 
\label{accept_rej_condition}
\end{equation}
This is know as metropolized independent sampler \cite{Liu,hastings,10.1214/aos/1176325750, roberts_rosenthal}. 

The fraction of the accepted propositions is another measure of the difference between $q_\theta$ and $p$. When $q_\theta=p$ acceptance ratio is 1. When two distributions are not identical some propositions can be rejected what leads to repetitions in the Markov chain. Therefore, even though propositions are independent from each other, a nonzero autocorrelation is possible. The range of those correlations is determined by the eigenvalues and eigenvectors  of the transition matrix of the Markov chain which are the subject of the next section. 

There are two obstacles limiting the scaling of that method to larger systems. First, the cost of generation of a single configuration of size $L^2=L\times L$ is equivalent to $L^2$ invocations of the neural network which by construction is proportional to a matrix-vector multiplication of length $L^2$. Hence, the total numerical cost of VAN algorithm scales with system size as $\sim L^6$. Second, the width of the neural network grows as $L^2$ with the system extent very quickly reaching sizes which are known to be difficult to train using current methodology. We address these issues in a separate publication, see Ref.~\cite{Bialas:2022qbs}.

\section{Analytic results for the eigenvalues of transition matrix}
\label{section: liu}

For the sake of completeness we include some results from Ref.\cite{Liu} and we use this opportunity to set up the notation. Although these results are known in the statistical Literature, they have not been employed in the studies of NMCMC algorithm so far. We consider two probability distributions: $q(\b s)$ and $p(\b s)$, defined on some set $X$. Distribution $p(\b s)$ is the target probability distribution. The Metropolised Independent Sampler (MIS) can be used to sample from $p$ by generating a proposal configuration $\b s_{k+1}$ from $q$ and accepting it with probability
\begin{equation}
    \mathcal{P}( \b s_k \rightarrow \b s_{k+1} ) = \min\left\{1, \frac{w(\b s_{k+1})}{w(\b s_{k})}\right\},
\end{equation}
where the importance ratio $w(\b s)$ is defined for each configuration $\b s \in X$ by
\begin{equation}
    w(\b s) = \frac{p(\b s)}{q(\b s)}.
    \label{eq:importance_ratio}
\end{equation}
Since $p(\b s), q(\b s) > 0$ it follows that $0 < w(\b s) < \infty$.\footnote{The last layer in the architecture of our neural networks is always the sigmoid function ensuring that $q(\b s) > 0$ and hence guaranteeing the ergodicity of the algorithm and finite values of $w(\b s)$.}
%always has a sIn principle $q(s)$ could become zero for some $s$, but then it would mean that this configuration would never be proposed as the trial configuration thus avoiding the problem of infinite $w(\b s)$}. 

The main result of Liu presented in Ref.\cite{Liu}  provides analytic formulae for the eigenvalues of the transition matrix of the resulting  Markov chain in terms of the importance ratios.  For a system with a finite number $M$ of elements in $X$ one can order the importance ratios $w(\b s)$  in an decreasing order (countable and continuous sets $X$ were also discussed in \cite{Liu}), so we adopt the labeling (which changes if $p$ or $q$ is modified) of configurations $\b s_i$ such that the corresponding importance ratios satisfy
\begin{equation}
\label{ordering_w}
    w(\b s_1) \ge w(\b s_2)  \ge \dots \ge w(\b s_M).
\end{equation} 
Direct diagonalization of the transition matrix of the Markov chain gives its eigenvalues $\lambda_k$, where
\begin{align}
    \lambda_k = \left\{ \begin{array}{cc}
    1 & \textrm{for } k = 0,\\
    \sum\limits_{i=k}^M q(\b s_i) \left( 1 - \frac{w(\b s_i)}{w(\b s_k)} \right) & \textrm{for } 0<k \le M-1
    \label{eigen_values}
     \end{array} \right.
\end{align}
The eigenvalue $\lambda_0=1$ corresponds to the stationary state and all remaining  eigenvalues $\lambda_{k>0}$ are nonnegative and smaller than 1. 
Due to the introduced ordering of importance ratios Eq.\eqref{ordering_w} and the fact that all $w$ are positive, it follows that $1>\lambda_1 > \lambda_2 > \ldots > \lambda_{M-1} > 0 $.

In the case of the Neural Markov Chain Monte Carlo algorithm, the sampling distribution $q(\b s)$ is encoded in the neural network and therefore depends on all the weights $\theta$ of the network, $q(\b s) = q_{\theta}(\b s)$. The aim of the training process is to ensure that $q_{\theta}(\b s)$ is as close to $p(\b s)$ as possible thus driving the importance ratios closer to 1 and hence decreasing the values of $\lambda_{k>0}$. This in turn translates into the reduction of the autocorrelation times as explained in more details in the next section. 

\begin{figure*}
\centering
\begin{tabular}{ll}
\includegraphics[ width=.4\textwidth]{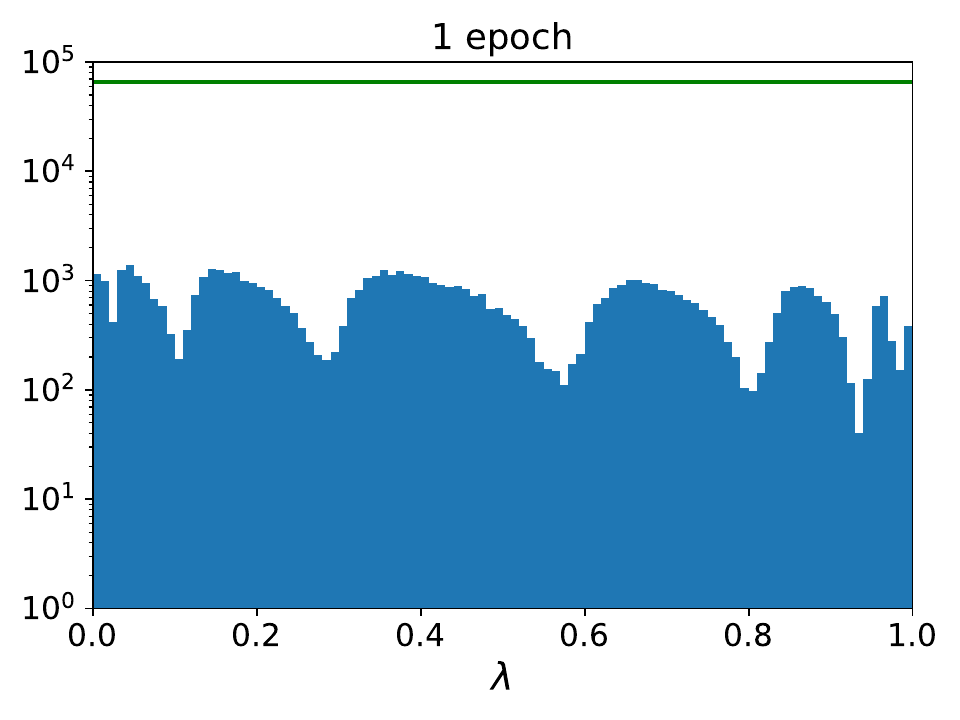} &
\includegraphics[ width=.4\textwidth]{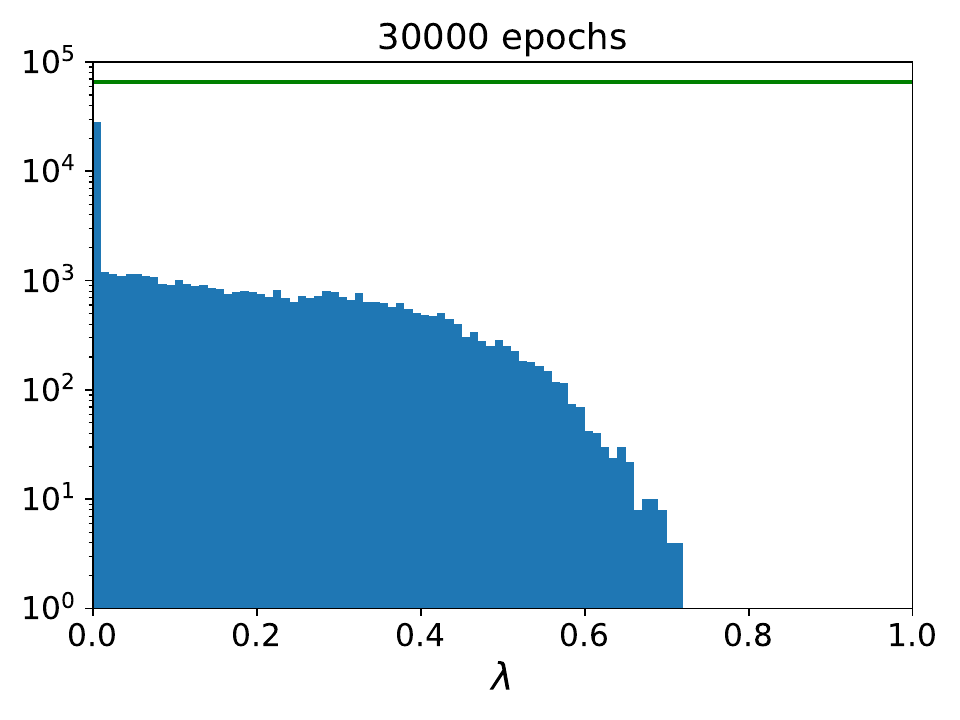} \\
\end{tabular}
\caption{Histograms of $M-1$ eigenvalues of transition matrix, $\lambda_{k>0}$, for system $4\times 4$. Training was performed using the $D_{\textrm{KL}}$ loss function. Left figure is for initial state of network, right is for fully trained network. Green line denotes $M-1=2^{16}-1=65535$ value.}
\label{lambdas_fig_4x4}
\end{figure*}

In Fig.~\ref{lambdas_fig_4x4} we show histograms of all eigenvalues of transition matrix smaller than 1 -- given by eq.~(\ref{eigen_values}) for Ising model of size $4 \times 4$ (so $2^{16}-1=65535$ eigenvalues in total). Temperature was chosen as $\beta=0.6$ and the $D_{\textrm{KL}}$ loss function was employed (see section \ref{section: loss functions}). Interval $(0,1)$ was divided into 100 bins.

The left histogram shows the situation when the network is at the initial state (1 epoch), right panel shows the histogram when the neural network is trained.
At the beginning of the simulation (1 epoch) the eigenvalues are distributed randomly with fluctuations coming from the random initial network weights. During the training process the eigenvalues decrease, as expected, since $q_\theta$ approaches $p$ distribution. The largest eigenvalue, $\lambda_1$ is around 0.75 when the network is fully trained.

\section{Autocorrelation times}
\label{sect_autocorr_times_sect}

In order to quantitatively discuss the algorithmic effectiveness we will estimate and compare the associated autocorrelation times. In practice there exist several ways to define the relevant autocorrelation time and they have been extensively studied in the Literature \cite{Sokal1997,Wolff:2003sm}. Here we review the main definitions and consider modifications needed to adapt them to the context of NMCMC algorithm. On one hand, one can focus on the transition matrix of the Markov chain and consider the analytically known or approximated autocorrelation time extracted from the eigenvalues of that matrix (see Section \ref{section: liu}). On the other hand, one can take the perspective of a particular observable of interest and estimate the autocorrelation time of that particular observable. In certain limiting situations all mentioned approaches should provide comparable estimates of the same, say longest, autocorrelation time. However, in the case of finite Markov chains and finite statistical precision, the autocorrelation times extracted along the two ways may vary by several orders of magnitude. We explain this effect in Section \ref{section: results small}, in the meantime we provide all needed definitions below. 

\subsection{Theoretical autocorrelation times }

The autocorrelation function for Markov chain is defined as
\begin{equation}
    \Gamma_\mathcal{O}(i,t)= \frac{1}{\var\, \mathcal{O} } E\left[\left(\mathcal{O}(\mathbf{s}_{i})-\bar{\mathcal{O}}\right) \left(\mathcal{O}(\mathbf{s}_{i+t})  - \bar{\mathcal{O}}\right)
    \right],
\label{autocorr_func}
\end{equation}
where the lower index enumerates consecutive configurations in the Markov chain. For stationary Markov chain  index $i$ can be  chosen arbitrarily and the result depends only on $t$, $\Gamma_\mathcal{O}(i,t)\equiv \Gamma_\mathcal{O}(t)$. Expectation value is calculated with respect to all realisations of the Markov chain and $\bar{\mathcal{O}}$   is the average of the observable $\mathcal{O}$. Normalisation is such that $\Gamma(0)=1$. 
This function can be  determined from the properties of the eigenvectors and eigenvalues of the Markov chain transition matrix and operator $\mathcal{O}$. In general autocorrelation function $\Gamma(t)$ for any observable $\mathcal{O}$ will be of the form:
\begin{multline}
\Gamma_{\mathcal{O}}(t) = \sum_{k>0} c_{\mathcal{O},k}^2\, \lambda_k^t = \sum_{k>0} c_{\mathcal{O},k}^2\, e^{-\frac{t}{\tau_k}},\\ \text{with } \tau_k = -(\log\lambda_k)^{-1}.
\end{multline}
The coefficients $ c_{\mathcal{O},k}$ can be calculated using the eigenvectors of transition matrix.  
One expects that for large separations $t$ the autocorrelation function decays exponentially:
\begin{equation}
    \Gamma_\mathcal{O}(t) \sim e^{-t/{\tOL}} \, \textrm{ for } t\rightarrow \infty.
\end{equation}
The $\tOL$ will be given by the largest eigenvalue $\lambda_k$ for  which $c_{\mathcal{O},k}>0$, and we can define 
\begin{equation}
\tL = \sup_{O} \tOL = -(\log\lambda_1)^{-1},  
\label{lambda_1}
\end{equation}
where $\sup_{O}$ is taken over all possible observables. 

In the extreme case when $w(\mathbf{s}) = 1$ for all $\mathbf{s}$, \textit{i.e.} $q=p$, it follows that $\lambda_{k>0} = 0$ and therefore $\tL = 0$. Note that we do not refer to any particular observable in defining $\tL$. It is a property of the Markov chain itself. Different observables will couple to the different modes of the Markov chain and will exhibit different leading autocorrelation times, which however will be always smaller or equal to $\tL$.

The most important measure of autocorrelations is the integrated autocorrelation time which enters as a factor into the definition of the statistical uncertainty of any value estimated from the Markov chain. For the observable $\mathcal{O}$  it is defined as \cite{Sokal1997}:
\begin{equation}
     \tOi =1+2\sum_{i=1}^{\infty}  \Gamma_\mathcal{O}(t).
\label{iat_def}
\end{equation}
In order to operate at the level of integrated autocorrelation times we can define "integrated version" of the $\tOL$:
\begin{equation}
    \tLOi \equiv 1+2\sum_{t=1}^{\infty} e^{-\frac{t}{\tOL}} 
    = \frac{1+ e^{-{1}/{\tOL}}}{1 - e^{-{1}/{\tOL}}}
\label{eq:tau_int_from_exp}    
\end{equation}
and similarly for $\tLi$
\begin{equation}
    \tLi = \frac{1+ e^{-{1}/{\tL}}}{1 - e^{-{1}/{\tL}}}.
\label{eq:tau_int_from_L}    
\end{equation}
Note that among those three theoretical autocorrelation times only $\tLi$ can be calculated without knowing $c_{\mathcal{O},k}$ coefficients.
What is more, this is possible only for very small systems where all states can be constructed explicitly. Calculating $c_{\mathcal{O},k}$ is beyond scope of this paper. Instead, we define below other estimators of integrated autocorrelation times which are calculable for systems of any size.

\subsection{Estimators of autocorrelation times for finite set of samples} 
\label{section_est_at}

Typically, in the context of NMCMC algorithm, in the training process one has access to relatively small sets of configurations generated from the distribution $q_{\theta}$. In this section we therefore relate the introduced theoretical autocorrelation times to their finite-statistics versions.

In practice we estimate $ \Gamma_\mathcal{O}(t)$ from the finite sample:
\begin{multline}
    \Gamma_\mathcal{O}(t) \approx \hat{\Gamma}_\mathcal{O}(t) = \\
    \frac{1}{\var\, \mathcal{O} }  \frac{1}{N_{sample}}\sum_{i=1}^{N_{sample}}\left( \mathcal{O}(\mathbf{s}_{i})-\bar{\mathcal{O}}\right) \left(\mathcal{O}(\mathbf{s}_{i+t}) - \bar{\mathcal{O}}\right),
\label{autocorr_func_practice}
\end{multline}
where $\bar{\mathcal{O}}$ and $\var\, \mathcal{O}$ are now mean and variance calculated using $N_{sample}$ states. In the following we mark all the  quantities that are estimated from finite sample with a hat sign. The autocorrelation function $\Gamma_\mathcal{O}(t)$ is always non-negative but its estimator $\hat\Gamma_\mathcal{O}(t)$  can become negative for some $t_{max}$ due to finite fluctuations. For the purpose of these studies we consider autocorrelation function for separations smaller than  $t_{max}$: $\hat\Gamma_\mathcal{O}(t), \ t< t_{max}$. 
This immediately gives us the estimator of integrated time Eq.~\eqref{iat_def}:
\begin{equation}
     \htOi =1+2\sum_{i=1}^{t_{max}}  \hat\Gamma_\mathcal{O}(t).
\label{iat_def_est}
\end{equation}

To estimate $\tLOi$ we fit the exponent to the autocorrelation function $\hat\Gamma$ at large $t$. In order to take into account also subleading autocorrelation time which can contribute at moderate $t$, we use two exponent function of the form:
\begin{equation}
    a_1 e^{-t/\tau_1}+ a_2 e^{-t/\tau_2}.
\label{fitted_function}
\end{equation}
We have checked that introducing more terms in Eq.~\eqref{fitted_function} does not change the value of largest $\tau_i$. We thus define $\htOF=\max(\tau_1,\tau_2)$ which describes the behaviour of $\hat\Gamma_\mathcal{O}(t)$ at large $t$ (but $t<t_{max}$). The integrated version of $\htOF$ is:
\begin{equation}
    \htOFi = 
     \frac{1+ e^{-{1}/{\htOF}}}{1 - e^{-{1}/{\htOF}}}.
\label{eq:tau_int_from_F}    
\end{equation}
The quantity $\htOFi$ given by this definition is an estimator of $\tLOi$ defined in Eq.~\eqref{eq:tau_int_from_exp}.

Also definition Eq.~\eqref{lambda_1} of $\tL$ has an estimator that is readily available and can be monitored during the training process of the neural network. The exact value of $\lambda_1$ is not available for most systems, but can be approximated by $\hat{\lambda}_1$ calculated from a finite sample of configurations (also called batch). Hence, we define the "batch version" of largest eigenvalue $\lambda_1$,
\begin{equation}
    \hat\lambda_1 \equiv \frac{1}{N_{\textrm{batch}}} \sum_{k=1}^{N_{\textrm{batch}}} \Big( 1 - \frac{w({\mathbf{s}_k})}{w_{\textrm{max on batch}}} \Big) \Big|_{\textrm{with } \mathbf{s}_k \textrm{ sampled from } q_{\theta}},
\end{equation}
and correspondingly the "batch version" of leading autocorrelation time $\tL$: 
\begin{equation}
  \htB = -(\log \hat\lambda_1)^{-1},
\end{equation}
which is an analogue of Eq.~\eqref{lambda_1}. Integrated version of this $\htB$ is:
\begin{equation}
    \htBi \equiv 
     \frac{1+ e^{-{1}/{\htB}}}{1 - e^{-{1}/{\htB}}}.
\label{eq:tau_int_from_B}    
\end{equation}

To conclude, we have defined here $\htOi$, $\htOFi$, $\htBi$ which are estimators of three versions of theoretical integrated autocorrelation times: $\tOi$, $\tLOi$ and $\tLi$.
In Appendix \ref{app. taus} we include a short discussion about the expected hierarchy of the values of these autocorrelation times.
In the following we choose the energy $H$, Eq.~\eqref{Ising_hamilt}, as observable $\mathcal{O}$. We have checked that the absolute value of the magnetization gives very similar results. We estimate $\hat{\Gamma}_\mathcal{O}(t)$ at the fixed stage of the training process and we generate a long, $\mathcal{O}(10^5)$, sequence of states. Subsequently we apply the condition Eq.~\eqref{accept_rej_condition} for consecutive states in this sequence and obtain a Markov chain whose stationary state is the Boltzmann distribution.

\section{Neural network loss functions}
\label{section: loss functions}

We explained in the previous section that the difference between the target probability distribution $p(\b s)$ and the sampling probability distribution $q_{\theta}(\b s)$ directly affects the autocorrelation times   via Eq.\eqref{eigen_values}. Several measures can be used to quantify the distance between two distributions, in the context of machine learning the most popular being the Kullback-Leibler divergence. Of course, in the ideal case, finding the global zero of any of the correctly defined measures will lead to finding such $\theta$ weights that $q_{\theta}(\b s) = p(\b s)$. However, for a finite size of the neural network and thus at a limited ability of the neural networks to model any probability function as well as for a limited number of training epochs, different measures may potentially be more efficient than others as they may lead to $q_{\theta}(\b s)$ that gives smaller autocorrelation times. Below we provide three definitions of loss functions which we will  apply to the case of two dimensional Ising model following sections.

\subsection{Kullback--Leibler (KL) divergence}
The most commonly used in the literature loss function is defined as:
\begin{align}
    D_\textrm{KL} (q_\theta | p) &= \sum_{k=1}^M q_\theta(\mathbf{s}_k) \, \log \left(\frac{q_\theta(\mathbf{s}_k)}{p(\mathbf{s}_k)}\right).
    \label{eq:KL_loss}
\end{align}
Substituting Eq.\eqref{eq_boltz_distr} into the last equation one obtains:
\begin{align}
    D_\textrm{KL} (q_\theta | p) &= \beta (F_q-F),
    \label{KL_loss_F}
\end{align}
where $F=- \frac{1}{\beta} \log Z$ is a true free energy and  
\begin{equation}
F_q= \frac{1}{\beta}\sum_{k=1}^M  q_\theta(\mathbf{s}_k) \left[\beta H(\mathbf{s}_k)+\log q_\theta(\mathbf{s}_k) \right]
\label{F_q_def}
\end{equation}
is variational free energy. Since $D_\textrm{KL}$ is a non-negative quantity the minimization of KL divergence is equivalent to minimization of the variational free energy.

As far as the importance ratios are concerned we can easily identify their presence in Eq.\eqref{eq:KL_loss}, since we can write
\begin{equation}
    D_\textrm{KL} (q_\theta | p) = - \sum_{k=1}^M q_\theta(\mathbf{s}_k) \, \log w(\mathbf{s}_k) = - \avg{\log w}_{ q_{\theta}}.
    \label{eq:DKL_loss}
\end{equation}
It should be clear from this reformulation that during the training process the neural network will try to improve $q_{\theta}$ for configurations for which the importance ratio is the smallest, i.e. for which $q_{\theta}(\mathbf{s}_k)$ is drastically overestimated with respect to $p(\mathbf{s}_k)$.

Calculating the gradient of the loss function 
\begin{equation}
    \diff{D_\textrm{KL} (q_\theta | p)}{\theta} =  -\diff*{\avg{\log w}_\qt}{\theta}
    \label{eq:DKL-grad}
\end{equation}
requires certain caution. The derivative in Eq.~\eqref{eq:DKL-grad} acts not only on the  expression being averaged but also on the distribution used for averaging. More generally 
\begin{multline}
\label{eq:avg-diff}
    \diff{}{\theta}\Big\langle S(\b s|\theta)\Big\rangle_\qt = 
  \diff*{\sum_{\b s} \qt(\b s) S(\b s|\theta)}{\theta}= \\
%   \sum_s \qt(s) \frac{1}{\qt(s)} \diff{ \qt(s)}{\theta} S(s|\theta) +\sum_s \qt(s)\diff{S(s|\theta)}{\theta}
% \end{equation}
% leading to 
% \begin{equation}
    % \diff{\avg{S(s|\theta)}_\qt}{\theta} = 
    \avg{\diff{\log \qt}{\theta} S}_\qt+ \avg{\diff{S}{\theta}}_\qt.
\end{multline}    
In the case of KL divergence, $S = \log q_\theta - \log p$, and 
\begin{equation}
    \avg{\diff{S}{\theta}}_\qt = 
    %\sum_s \qt(s) \frac{1}{\qt(s)}\diff{\qt(s)}{\theta} =
    \diff*{\sum_s \qt(s)}{\theta}=0.
\end{equation}
So finally 
\begin{equation}
    \diff{D_\textrm{KL} (q_\theta | p)}{\theta} 
    =\avg{ \diff{ \log q_\theta}{\theta}  (\log q_\theta - \log p)}_{q_\theta}.
    \label{DKL_grad}
\end{equation}
As was noticed in Ref.~\cite{2019PhRvL.122h0602W}, Eq.~\eqref{DKL_grad} can be interpreted as a REINFORCE algorithm \cite{Reinforce_book} with score function $-S$.

The KL divergence can be seen as a special case of a more general loss function known as $\alpha$-divergence \cite{minkadivergence}. It has been identified that such loss function can exhibit various behaviours when the target distribution $p$ has multiple modes, either the model tends to get trapped in one of the minimum leaving other minima unexplored (mode-seeking behavior) or it has the power of travel between different modes (mass-covering behavior). The KL divergence can be shown to belong to the mode-seeking category, however the severity of that problem may depend on various parameters, for example $\beta$, and thus should be checked in each case separately. For detailed discussion of this topic in context of Normalizing Flows see Ref.~\cite{Hackett:2021idh}.

\subsection{Largest Eigenvalue of transition matrix $\lambda_1$}

An alternative way to measure the difference between $q_{\theta}$ and $p$ is to use directly the expression for the largest eigenvalue of the associated Markov chain transition matrix provided by  Eq.~\eqref{eigen_values} 
\begin{multline}
    D_{\lambda_1} (q_\theta | p) \equiv \lambda_1 = \sum_{k=1}^M q_\theta(\mathbf{s}_k) \left( 1 - \frac{w(\b s_k)}{w(\b s_1)} \right)
    =\\ \avg{ 1 - \frac{w}{w(\b s_1)}}_{q_\theta}.
    \label{eq:LAT_loss}
\end{multline}
Because $\avg{w}_{q_\theta}=1$ 
\begin{equation}
    D_{\lambda_1} (q_\theta | p) = 1-\frac{1}{w(\b s_1)}.
    \label{eq:lambda_1}
\end{equation}
We cannot however use this form without knowing the the overall normalization factor, i.e. partition function $Z$.  In \eqref{eq:LAT_loss} $p$ enters only as a  ratio and so the partition function $Z$ is not needed to calculate the eigenvalue.
% where we have used the facts that $\sum_k q_{\theta}(\mathbf{s}_k) = 1$ and $\sum_k q_{\theta}(\mathbf{s}_k)w(\b{s}_k) = \sum_k p_{\theta}(\mathbf{s}_k) = 1$.
As already stated, $\lambda_{1} \ge 0$ and it vanishes only when $q_{\theta} \equiv p$. 

Although this loss function has the same minimum as the KL divergence, the contributions coming from particular configurations are different, and hence during minimization $q_{\theta}$ will be brought closer to $p$ for a different subset of configurations first. %We see that the main contribution to $\dlam$ come from configurations which have the largest importance ratios (configurations for which $q_{\theta}(\mathbf{s}_k)$ is drastically underestimated). 

Using Eq.~\eqref{eq:avg-diff} one can obtain the gradient of $D_{\lambda_1}$,
\begin{align}
    \diff{D_{\lambda_1} (q_\theta | p)}{\theta} =& \avg{\diff{\log q_\theta}{\theta}}_{q_\theta}-\avg{\frac{w}{w(\b s_1)}\diff{\log q_\theta(\mathbf{s}_1)}{\theta}}_{q_\theta}.
    \label{grad_Dlam1}
\end{align}
It is easy to see that this gradient does not vanish when $p=q_\theta$, because $D_{\lambda_1}$ has a cusp at this point.

Alternatively, one can follow the REINFORCE algorithm and use the following quantity in backward propagation,
\begin{equation}
\avg{ \diff{ \log q_\theta}{\theta}  \left( 1 - \frac{w}{w(\b s_1)}\right) }_{q_\theta},
\label{reinforce_lam_1}
\end{equation}
instead of the derivative Eq.~\eqref{grad_Dlam1}. This quantity vanishes at $p=q_\theta$. Note that, in contrast to $D_\textrm{KL}$, these two approaches are different. Our numerical studies showed that the approach given by Eq.~\eqref{reinforce_lam_1} results in a much more efficient training for the neural network. The detailed comparison of these two approaches and their efficiencies is beyond scope of this paper. In what follows we show results only for the REINFORCE approach.

\subsection{Effective Sample Size (ESS)}
Following Ref.\cite{Liu,2021arXiv210108176A} we define yet another possible loss function, which for the sake of completeness we compare with the two previously defined,
\begin{equation}
    ESS= \frac{\avg{w}_{q_\theta}^2}{\avg{w^2}_{q_\theta}}.
\label{ESS_definition}
\end{equation}
The quantity is normalized such that $ESS\in [0,1]$ and for perfectly matched distributions one has $ESS=1$. So we define the loss function as $1-ESS$
\begin{equation}
D_\textrm{ESS} (q_\theta | p) = 1 -\frac{\avg{w}_{q_\theta}^2}{\avg{w^2}_{q_\theta}}
    \label{eq:ESS_loss} = \frac{\avg{w^2}_{q_\theta}-\avg{w}_{q_\theta}^2}{\avg{w^2}_{q_\theta}} =  \frac{\var_{q_\theta}{w}}{\avg{w^2}_{q_\theta}}.
\end{equation}
Again, the partition function $Z$ is not needed to calculate $ESS$.

This loss function cannot be expressed as a weigthed average so the REINFORCE algorithm is not available. We differentiate Eq.~\eqref{eq:ESS_loss} obtaining:
\begin{equation}
    \diff{D_\textrm{ESS} (q_\theta | p)}{\theta} = 
    -\frac{\avg{w}_\qt^2}{\avg{w^2}_\qt^2} \avg{\diff{\log \qt}{\theta}\; w^2}_\qt.
\end{equation}

\subsection{Batch versions of the loss functions}

For all but very small systems, the sum over all configurations $\sum_{i=k}^M(\ldots)$ is prohibitively large to perform exactly. In the training process one disposes of a set of configurations $\mathbf{s}$ drawn from the current probability distribution $q_{\theta}(\mathbf{s})$. Hence, all the three loss functions defined above can be expressed as batch averages. To this goal we do the following replacements in Eqs.~\eqref{eq:DKL_loss}, \eqref{eq:LAT_loss} and \eqref{eq:ESS_loss}:
\begin{equation}
 \avg{\ldots}_{q_\theta} =\sum_{i=1}^M q_\theta(\mathbf{s}_i)(\ldots) \rightarrow \frac{1}{N_{batch}} \sum_{i=1}^{N_{batch}} (\ldots).
\label{eq:qtheta_av_batch_averaging}
\end{equation}

For the KL divergence we thus have,
\begin{multline}
    D_\textrm{KL} (q_\theta | p) \approx  \hat{D}_\textrm{KL} (q_\theta | p) \equiv \\ - \frac{1}{N_{\textrm{batch}}} \sum_{k=1}^{N_{\textrm{batch}}} \log w(\mathbf{s}_k) \Big|_{\textrm{with } \mathbf{s}_k \textrm{ sampled from } q_{\theta}} = \\ - \langle \log w \rangle_{\textrm{batch average}},
\end{multline}
where we used the hat symbol to denote quantity calculated from the batch of configurations. In the case of $D_{\lambda_1} (q_\theta | p)$ we also need to modify the value of $w_1$ since in practice its value may be unknown. We thus say
\begin{multline}
%\begin{split}
    D_{\lambda_1} (q_\theta | p) \approx \hat{D}_{\lambda_1} (q_\theta | p)\equiv  \\ \frac{1}{N_{\textrm{batch}}} \sum_{k=1}^{N_{\textrm{batch}}} \Big( 1 - \frac{w({\mathbf{s}_k})}{w_{\textrm{max on batch}}} \Big) \Big|_{\textrm{with } \mathbf{s}_k \textrm{ sampled from } q_{\theta}} \\
    = \langle 1 - \frac{w}{w_{\textrm{max on batch}}} \rangle_{\textrm{batch average}}
    \label{hat_lambda_1}
%\end{split}
\end{multline}
It should be clear that $w(\b s_1) \ge w_{\textrm{max on batch}}$. This however does not necessarily entails that $\hat{D}_{\lambda_1} (q_\theta | p)$ is smaller than $D_{\lambda_1} (q_\theta | p)$ as the relation Eq.\eqref{eq:lambda_1}  need not hold for the finite sample.

\section{Small lattice}
\label{section: results small}

In Section \ref{section: liu} we recalled analytical results concerning NMCMC algorithms, in particular explicit expressions for the eigenvalues of the transition matrix. The evaluation of these formulae is only possible when the system size is small and one can visit all configurations $\b s \in X$. The aim of this section is to study a system small enough, we choose the lattice of size $4 \times 4$, and investigate the differences between the exact analytic results and their approximated batch versions. For systems of physically reasonable sizes only the latter are available and therefore it is important to understand their properties. 

\subsection{Neural network architecture}

In this work we employ the simplest implementation of an autoregressive neural network with dense layers. We use one hidden layer and the width of all layers is equal to the number of spins on the lattice. As activation functions we use PReLU in between layers and sigmoid for output layer. Adam optimizer is used. The $\beta$-annealing is used for training, except for Figs.~\ref{lambdas_fig_4x4},
\ref{fig:dynamics logqlogp}, \ref{fig:dynamics logw}. We demonstrate in the remaining of this section that such an architecture is sufficient for our purposes. We checked that deeper neural networks do not provide improvements in the final results, in particular no significant change can be seen in the autocorrelation times. 

\subsection{Dynamics of the neural network during training}

\begin{figure*}
\centering
\begin{tabular}{llll}
{\large $\beta=0.6$:}\\
\includegraphics[ width=.22\textwidth]{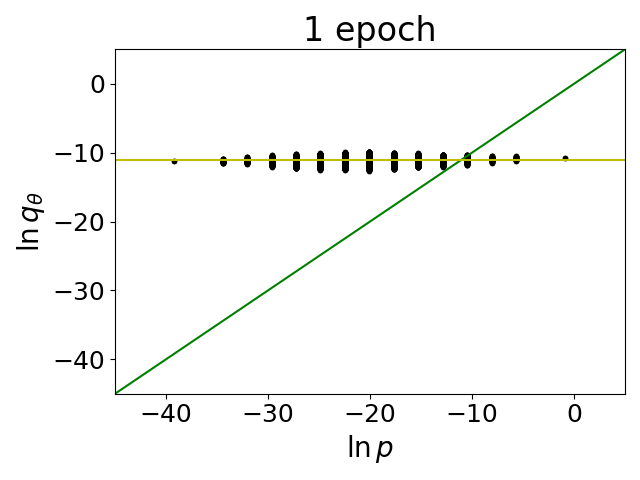} &
\includegraphics[ width=.22\textwidth]{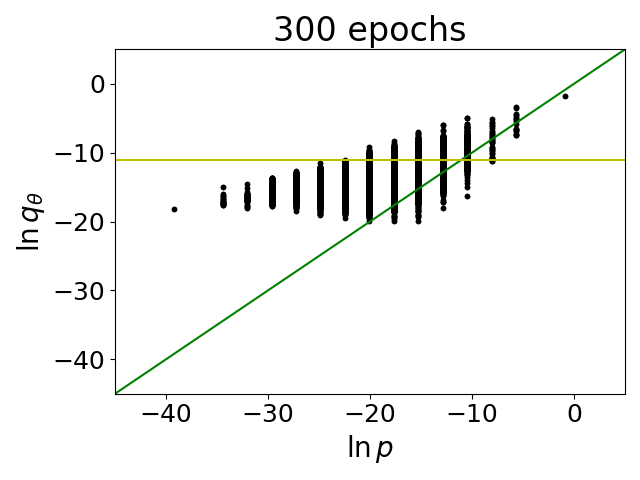} &
\includegraphics[ width=.22\textwidth]{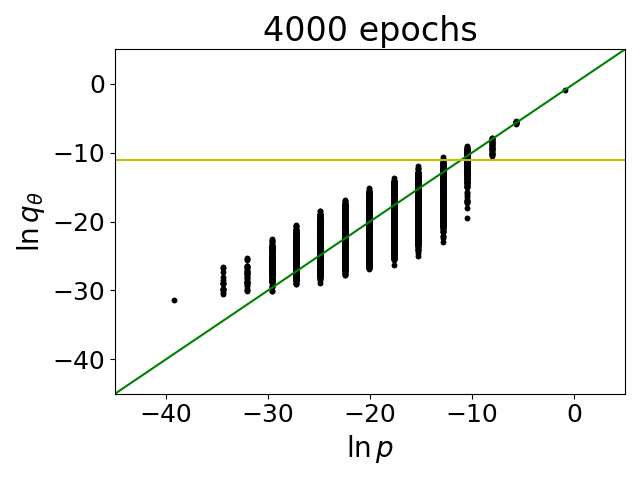} &
\includegraphics[ width=.22\textwidth]{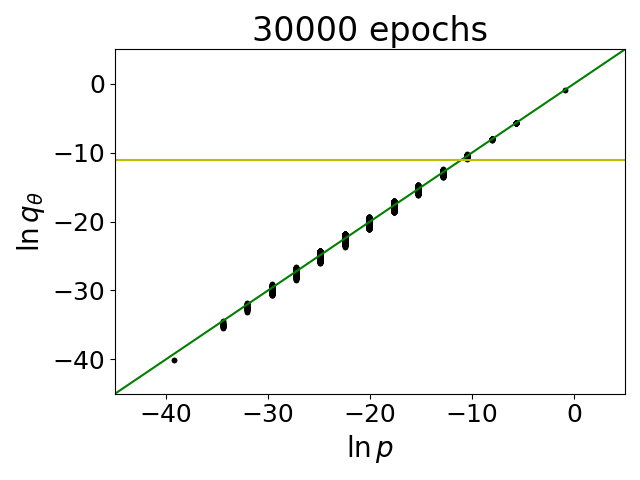} \\
{\large $\beta=0.3$:}\\
\includegraphics[ width=.22\textwidth]{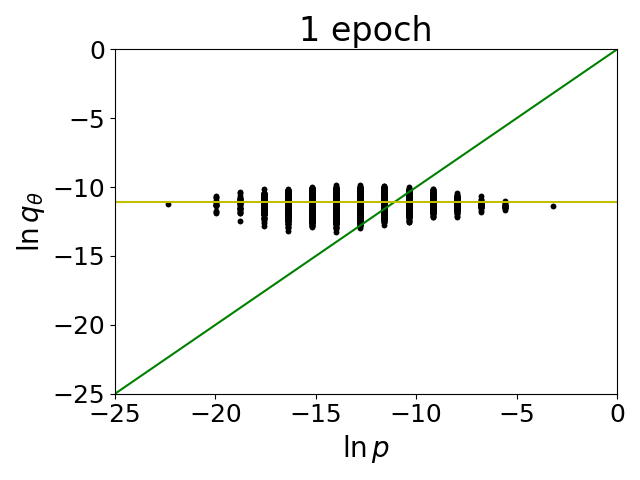} &
\includegraphics[ width=.22\textwidth]{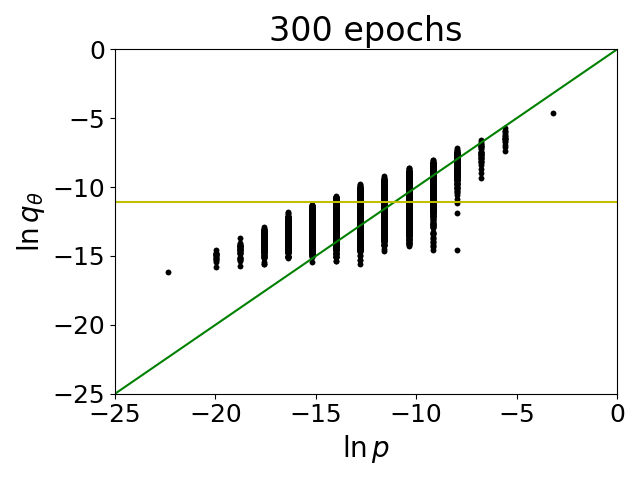} &
\includegraphics[ width=.22\textwidth]{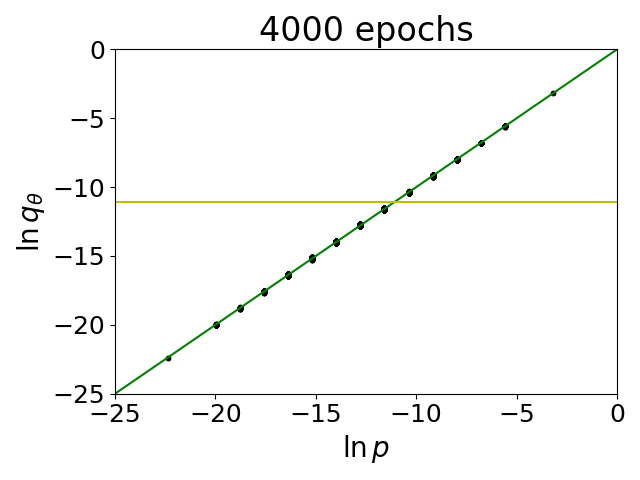} &
\includegraphics[ width=.22\textwidth]{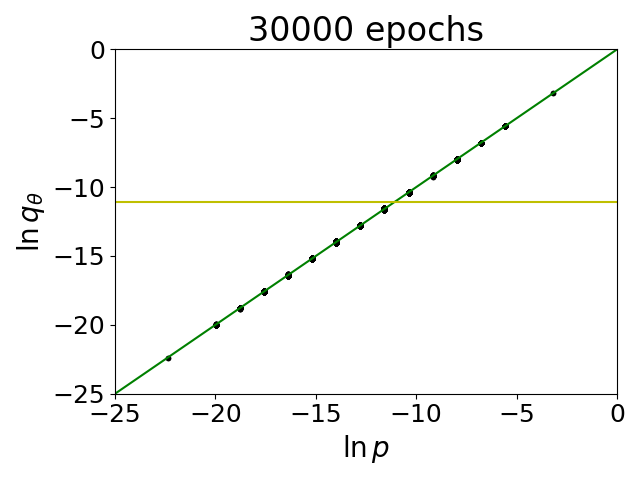} \\
\end{tabular}
\caption{Dynamics of the neural network training using the KL divergence loss function Eq.~\eqref{eq:DKL_loss}.  For two inverse temperatures: $\beta=0.6$ (upper row) and $\beta=0.3$ (lower row) we provide snapshots after 1, 300, 4000 and 30000 epochs of $\log q_\theta$ vs $\log p$. It is evident from the plots that the training is more efficient for smaller values of $\beta$: dots representing states position themselves on $\log q_\theta =\log p$ line (green) much faster. The yellow horizontal line shows a uniform probability distribution of $p(\mathbf{s}) = 2^{-16}$.}
\label{fig:dynamics logqlogp}
\end{figure*}

The small size of the $4 \times 4$ Ising model allows us to monitor the dynamics of the neural network during the training process. All loss functions have a common, unique minimum where the probabilities provided by the neural network, $\log q_\theta $, are equal to the Boltzmann probabilities, $\log p$. By plotting the former against the latter at several values of the training time we can monitor how the differences between the two evolve during the training. In the presented case we used the KL divergence loss function Eq.~\eqref{eq:DKL_loss}. Other loss functions show a similar quantitative behaviour, with  differences at the intermediate stages of the training associated to the different weighting of contributions from particular configurations. At the initial moment the weights in the neural network are random and all probabilities are approximately equal. They form a horizontally elongated cloud along the yellow line shown in Fig.~\ref{fig:dynamics logqlogp}. The yellow line corresponds to a uniform probability distribution, $q_{\theta}(\b s) = 2^{-16}$. When $\log q_\theta  = \log p$ the data points form a bisecting line, as can be seen in the last stages of the training shown in the most right hand side plots in Fig.~\ref{fig:dynamics logqlogp}. The two plots in-between are the most interesting ones as they capture the way the algorithm improves on $q_{\theta}$. As expected, first the configurations which have the largest probabilities $\log p$ (upper right corner of the figures) are tuned to the desired values and position themselves on the straight line. Once they have the correct values the training improves the approximation of less probable configurations. Please note that at the same time the approximated probabilities for configurations which are very rare are also improved, although it is very unlikely that they have appeared in any batch. This is due to the generalisation properties of the network. Appendix \ref{app. dynamics} contains additional plots showing the dynamics of the training.

\subsection{Variational free energy}

\begin{figure}
\includegraphics[width=.48\textwidth]{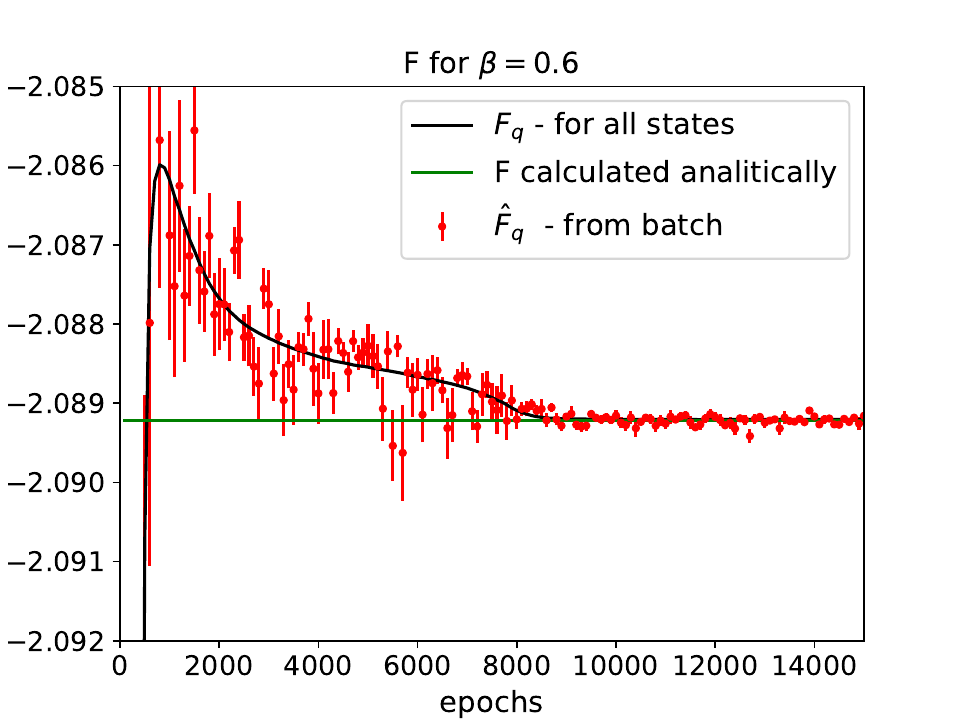} 
\includegraphics[ width=.48\textwidth]{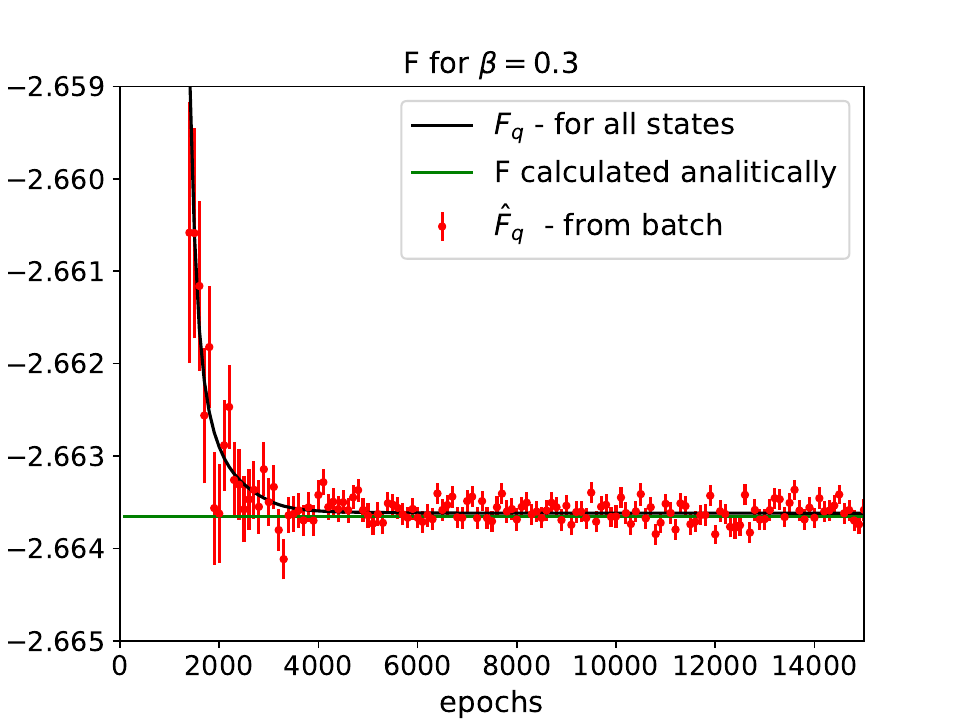} 
\caption{Free energy $F$ for system of size $4\times 4$ at $\beta=0.6$ and $\beta=0.3$. We show value estimated in the batch $\hat F_q$ (red dots) and value $F_q$ calculated using all states (black curve). They are compared with analytical result for $F$ (green horizontal line). It has to be emphasized that the red data obtained from the batch averages follows closely the full estimate of the free energy plotted with the black line.
}
\label{F_fig_4x4}
\end{figure}

Now we turn our attention to the variational estimation of the free energy given by Eq.~\eqref{F_q_def}. This is a particular observable for which we search for a variational approximation from above using the setup we have described in previous section. Note, that the results presented below are not outcomes from a Monte Carlo simulation. The discussion below will serve two purposes. On one hand, we demonstrate how accurately the correct probability distribution can be reproduced from the point of view of a physically relevant observable. On the second hand, this benchmark observable can be used to select the simplest neural network architecture which will be sufficient for our purposes. In Fig.\ref{F_fig_4x4} we show the free energy evaluated in three different ways as a function of the training epoch at $\beta=0.6$ (upper panel) and $\beta=0.6$ (lower panel). We used a small neural network architecture -- 2 dense layers with  autoregressive connections. Each layer contains $16$ (i.e. number of spins) neurons. We use the $Z_2$ symmetry of the system (see also Section \ref{section: symmetries}). 
Different data points in Fig.~\ref{F_fig_4x4} correspond to: Eq.~\eqref{F_q_def} which is the full $F_q$ (black curve), the batch estimate $\hat F_q$ which is Eq.~\eqref{F_q_def} with replacement Eq.~\eqref{eq:qtheta_av_batch_averaging} (red dots) and the value of $F$ calculated analytically (green horizontal line) -- see Ref.~\cite{PhysRev.185.832} or \cite{2020PhRvE.101b3304N} for formulas. The convergence of $F_q$ and $\hat F_q$ to the exact value $F$ confirms that such a very small neural network architecture was enough for this system size. Additionally, we can remark that two ways of calculating the free energy, $F_q$ and $\hat F_q$, agree very well. Note also that at early epochs $F_q$ is smaller than true $F$ because $\beta$-annealing was used. A similar test performed at other values of $\beta$ confirmed that the neural network architecture is sufficient. When considering larger system size we will keep the number of layers equal to 2 and only adjust their width to match the increasing number of spins.

\subsection{Autocorrelation times}

\begin{figure}
\includegraphics[ width=.48\textwidth]{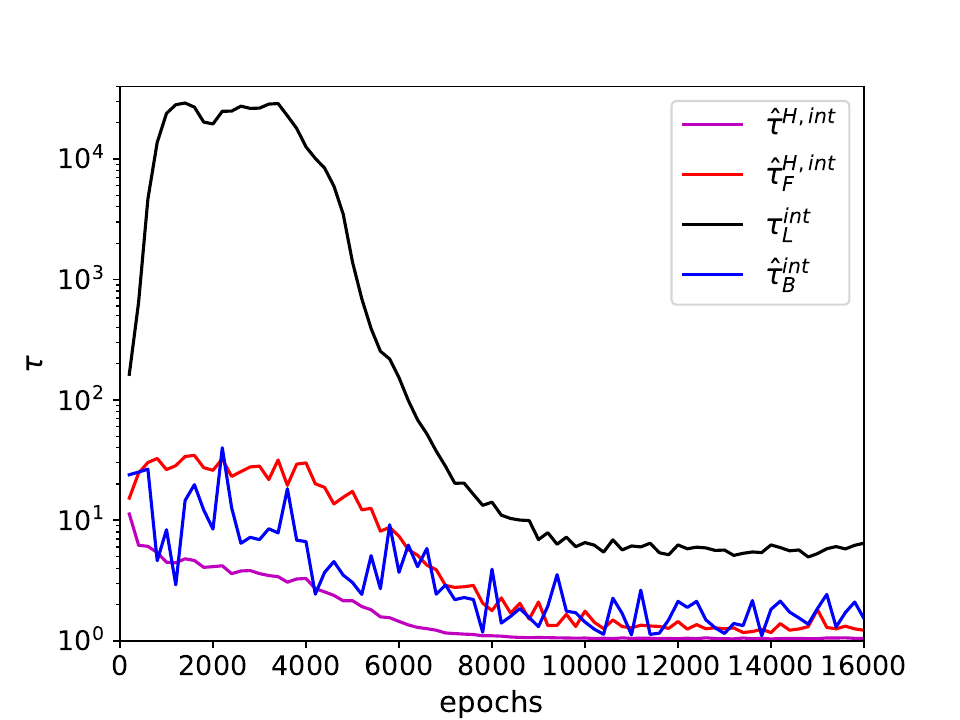} 
\includegraphics[ width=.48\textwidth]{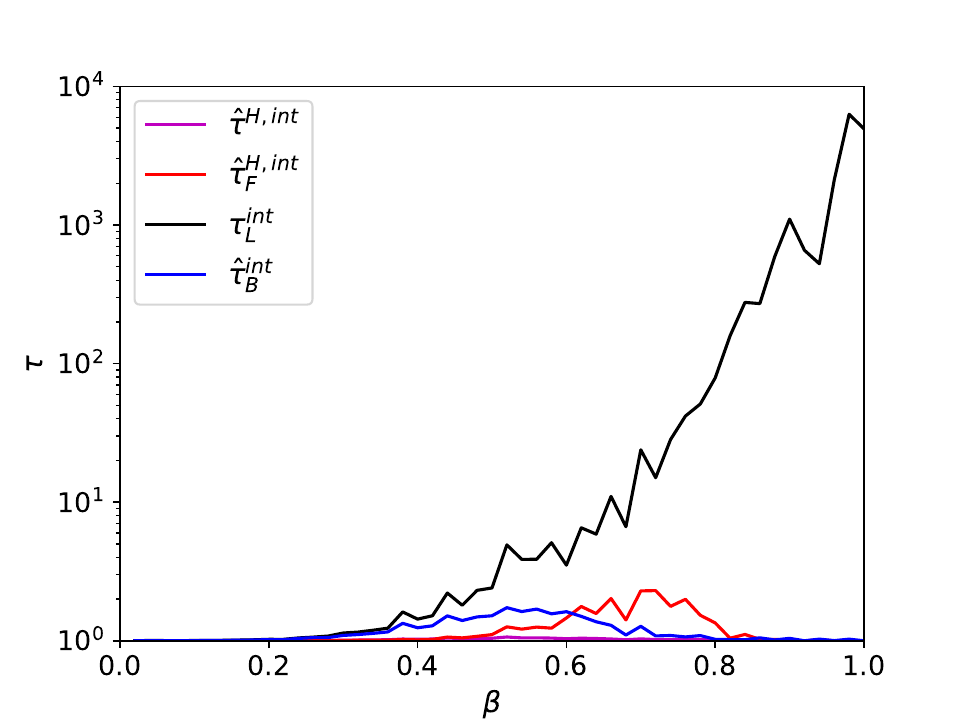}
\caption{Four versions of (integrated) autocorrelation time (defined in section \ref{sect_autocorr_times_sect}) for system of size $4\times 4$. Top panel: autocorrelation times during the training process for $\beta=0.6$. Bottom panel: autocorrelation times as function of $\beta$, they were measured at epoch 15000, when networks were fully trained. The system size $4 \times 4$ allows us to compare the theoretical integrated autocorrelation time with the other definitions. The behaviour of autocorrelation times for a large system size $16 \times 16$ is shown in Fig.\ref{taus_fig_16x16} in which case theoretical autocorrelation time $\tLi$ is not available.
}
\label{taus_fig_4x4}
\end{figure}

The next quantity which we can discuss at this small system size are the different integrated autocorrelation times defined in section \ref{sect_autocorr_times_sect} by equations ~\eqref{eq:tau_int_from_L}, \eqref{iat_def_est}, \eqref{eq:tau_int_from_F} and \eqref{eq:tau_int_from_B}.
We start by showing how the autocorrelation time changes during the training process. The top panel of Fig.~\ref{taus_fig_4x4} shows the evolution at $\beta=0.6$ as a function of training epochs. The most striking is the black curve corresponding to the theoretical integrated autocorrelation time $\tLi$, Eq.~\eqref{eq:tau_int_from_L}. It provides the reference scale for the worst possible behaviour. It is interesting that after a period of very pronounced values larger than $10^4$ it decreases and stabilizes around 10 when the training is finished. The transition time coincides with the time at which the variational free energy converges towards the exact value, see Fig.~\ref{F_fig_4x4}, i.e. around 8000 epochs. Similar behaviour can be seen for the other estimators, they all show a correlated drop in this period of training. 

The large discrepancy between $\tLi$ and the other three curves has two sources. The large value of $\tLi$ is driven by some underestimated configuration which  has a very small probability to appear in the batch, therefore, it will not contribute to $\tB$. The configurations which contribute to $\tB$ are those which are generated in the batch and these are the same configurations which are used to improve $q_{\theta}$. Hence, $\tB$ is estimated from configurations whose importance ratios are closer to one, leading to smaller autocorrelation time. The other source is related to the fact that $\tF$ and $\tI$ are calculated for a specific observable, in our case the energy, and hence may couple differently to different modes of the Markov chain. The discrepancy with $\tLi$ signals that the contribution of the slowest mode is small in the case of such physical observables as energy or magnetization. As far as the statistical uncertainties of such physical observables are concerned, it will be $\tI$ and $\tF$ which matter. It is worth noticing that $\tB$ is very close to $\tF$ and $\tI$, hence it provides a reasonable estimator of the autocorrelation time in the Markov chain without making reference to any specific observable. The increased fluctuations of $\tB$ are present because at each epoch new states in the batch are drawn and different importance ratios appear.

In lower panel of Fig.~\ref{taus_fig_4x4} we show the dependence of the various definitions of the autocorrelation time estimated after 15000 epochs of training as a function of $\beta$. From the traditional Metropolis algorithm we expect a sharp maximum around the ordered/disordered phase cross-over inverse temperature around $\beta \approx 0.4$, however no rise of autocorrelation time is observed in this region. What we see instead is a strong rise of $\tLi$ which signifies that there are configurations for which $q_{\theta}$ provides an increasingly badly underestimated approximation to the target probability. Such configurations have a very small probability of being generated and so they did not appear in any  batch hence do not contribute to any other autocorrelation time, $\tI$, $\tF$ and also $\tB$ are always of the order of 1.0.

\section{Medium lattice: loss functions and critical slowing down}
\label{section: results large}
 
We now move to describe results from larger lattices. We study the Ising model on a lattice with extent $L=16$. For this linear extent the system is described by $2^{256}$ configurations and the Liu's analytic solution, Eq.~\eqref{eigen_values}, would require the estimation of $q_{\theta}$ for all these configurations which cannot be done directly. Instead, one has to rely on quantities which can be estimated as batch averages. We showed how the two, the exact one and the batch-approximated one, correlate in a smaller system in the previous section. 

We consider different loss functions and study the behaviour of several observables as a function of training epochs to visualise some aspects of the neural network training process. Again we keep the simplest neural network architecture, with two dense layers of width equal to the number of spins on the lattice: 256. We checked that deeper architectures do not provide an improvement of the results, but they increase the necessary number of epochs in the training phase.

\subsection{Comparison of loss functions}

\begin{figure*}
\centering
\begin{tabular}{ll}
\includegraphics[ width=.45\textwidth]{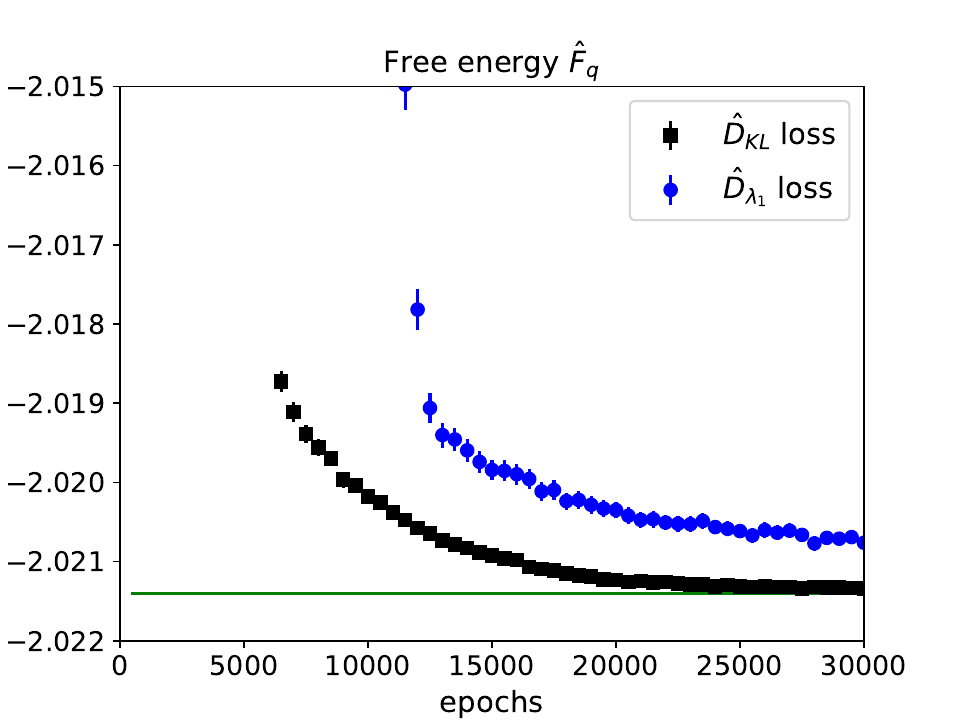} &
\includegraphics[ width=.45\textwidth]{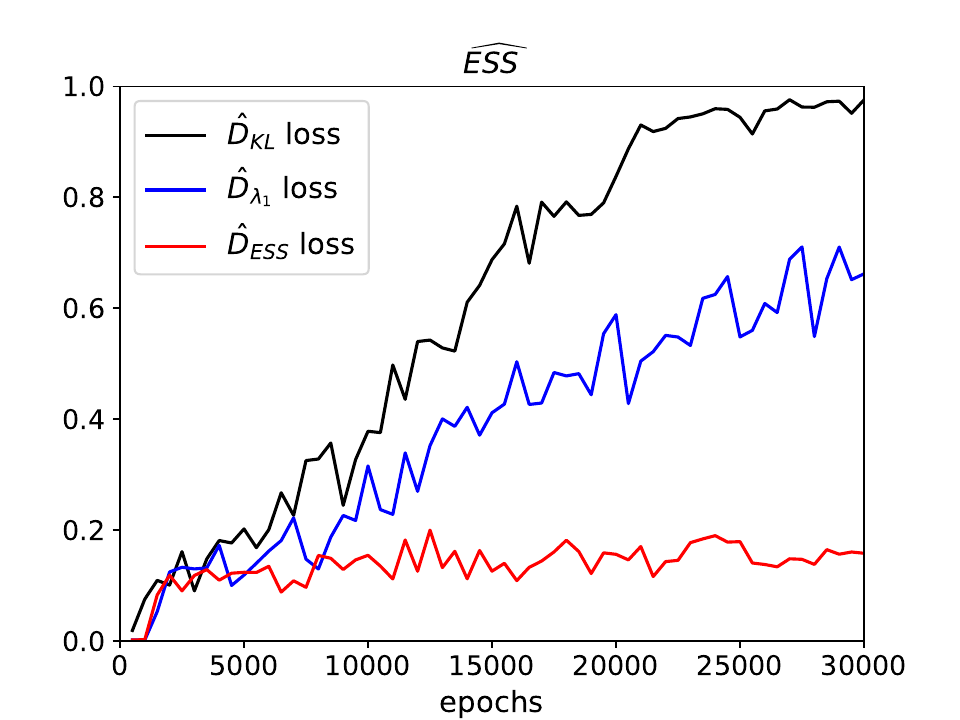}\\
\includegraphics[ width=.45\textwidth]{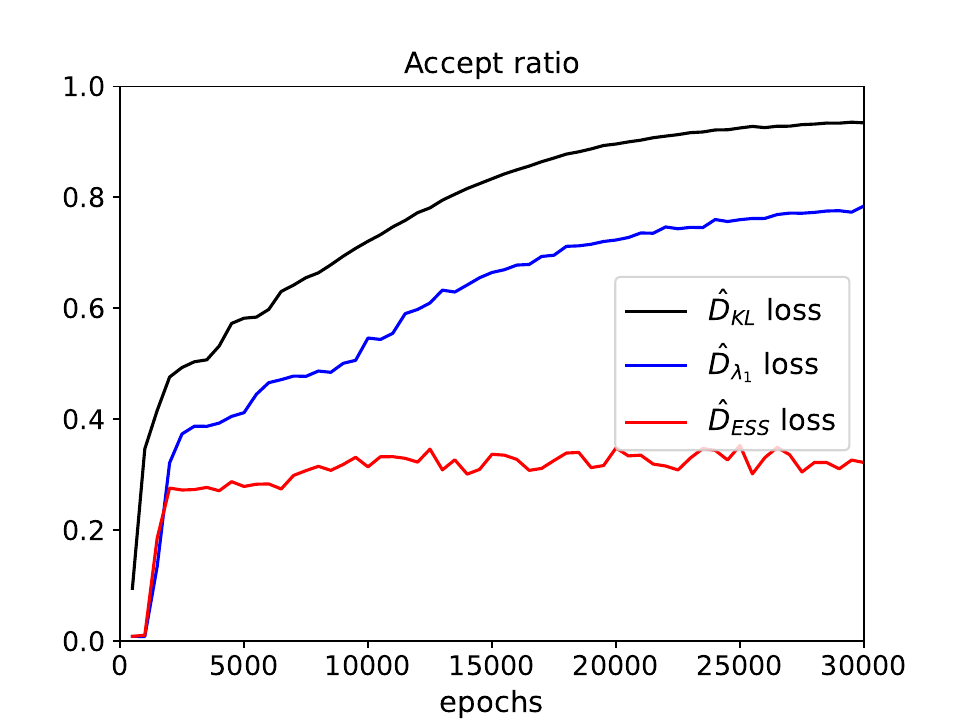}&
\includegraphics[ width=.45\textwidth]{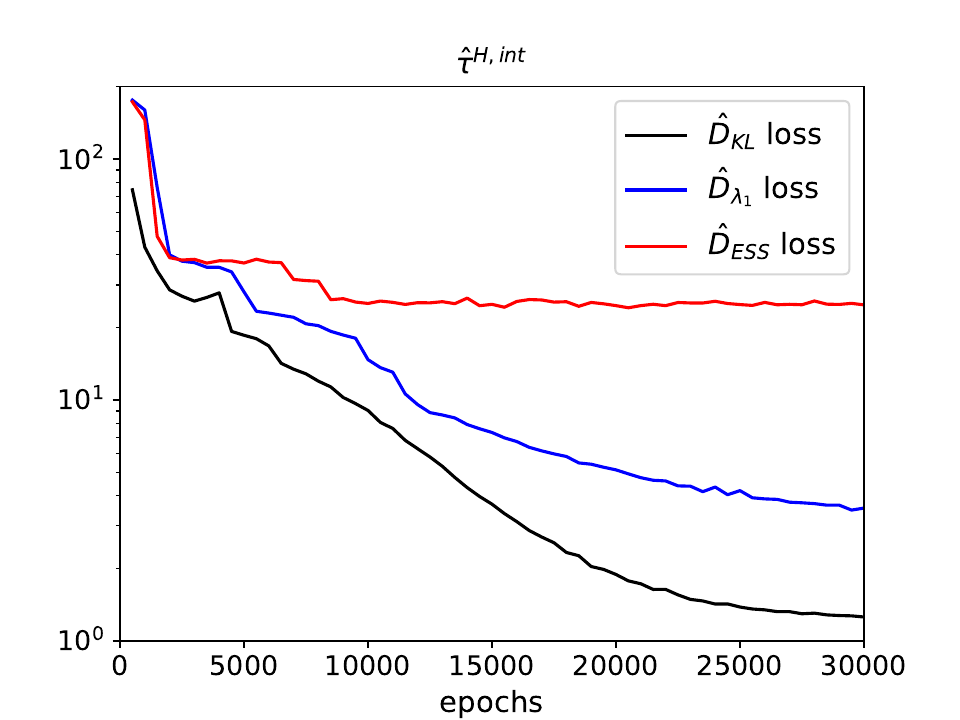}\\
\includegraphics[ width=.45\textwidth]{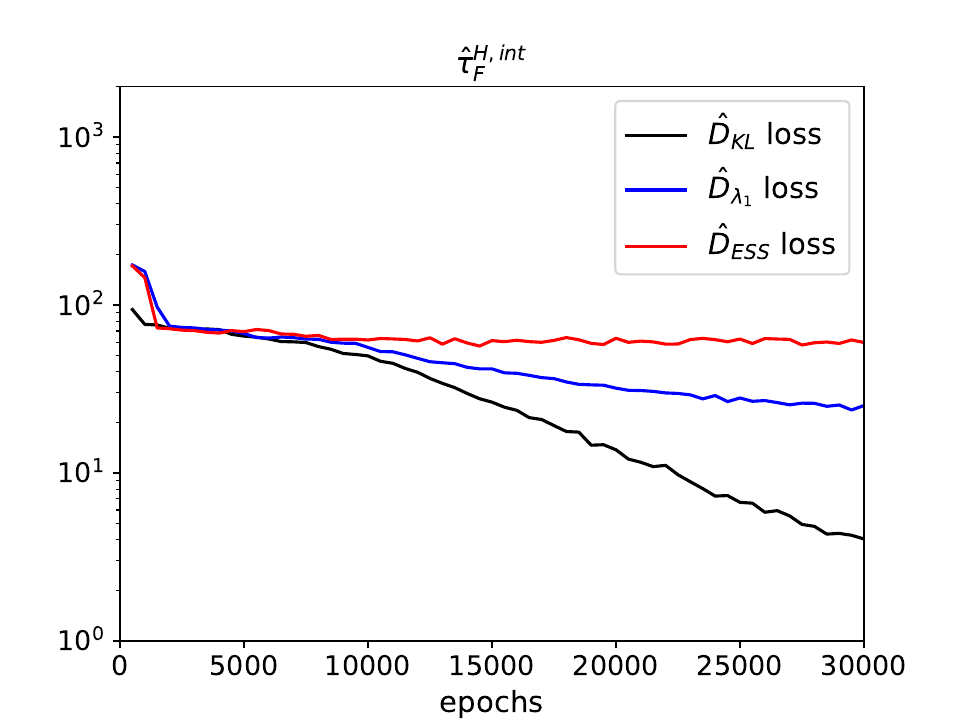}&
\includegraphics[ width=.45\textwidth]{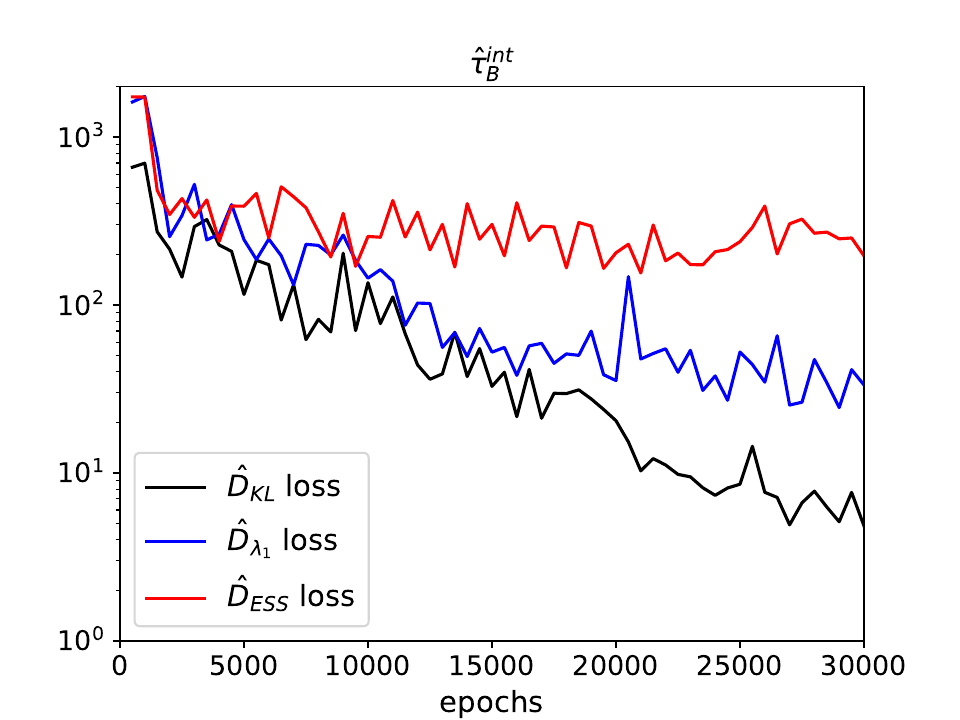}
\end{tabular}
\caption{Learning process for $16\times 16$ system at $\beta=0.6$ using three loss functions. Different observables have been shown as functions of training epochs. Functions shown on the plots are averages of 16 learning process in order to exclude effects of random net initialization. The free energy for training with $\dess$ did not converge and is out of the scale of the figure.}
\label{loss_funct_comparison}
\end{figure*}

\begin{figure*}
\centering
\begin{tabular}{ll}
\includegraphics[ width=.45\textwidth]{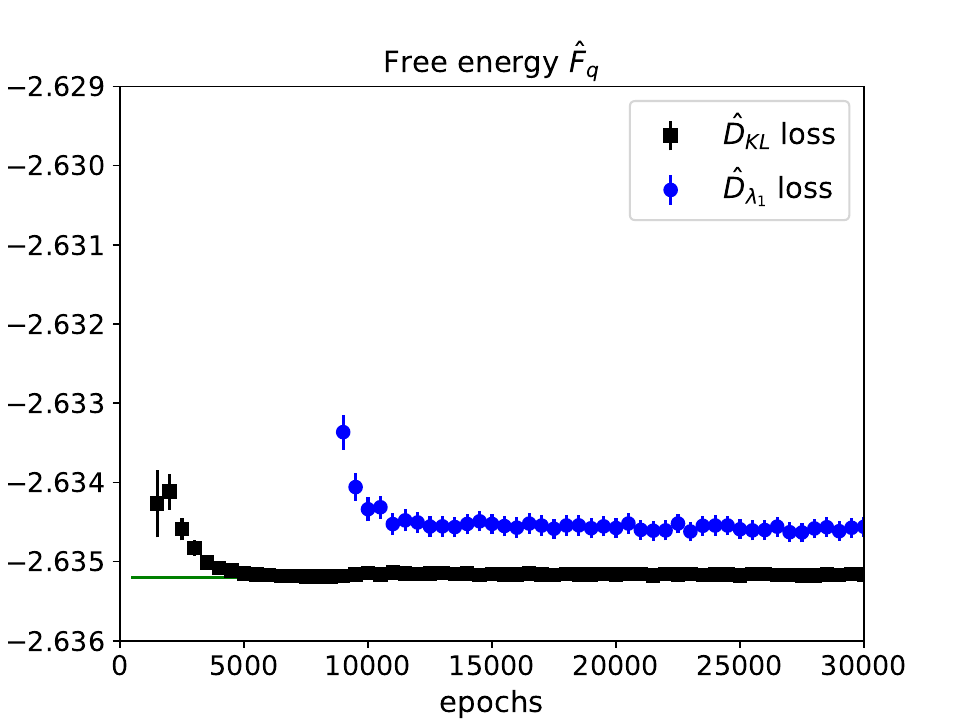} &
\includegraphics[ width=.45\textwidth]{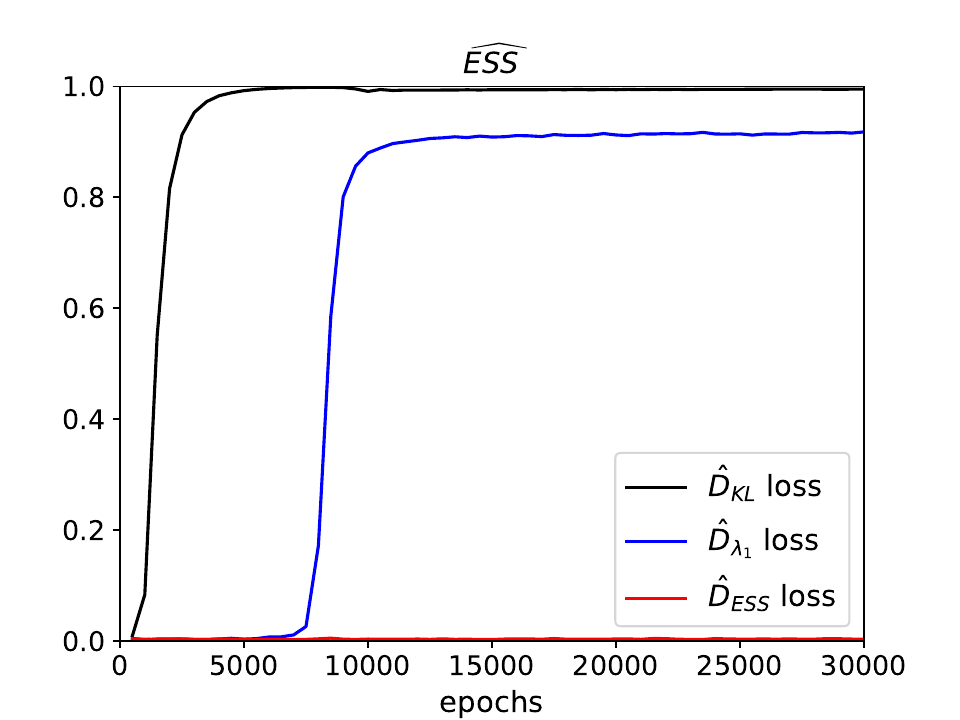}\\
\includegraphics[ width=.45\textwidth]{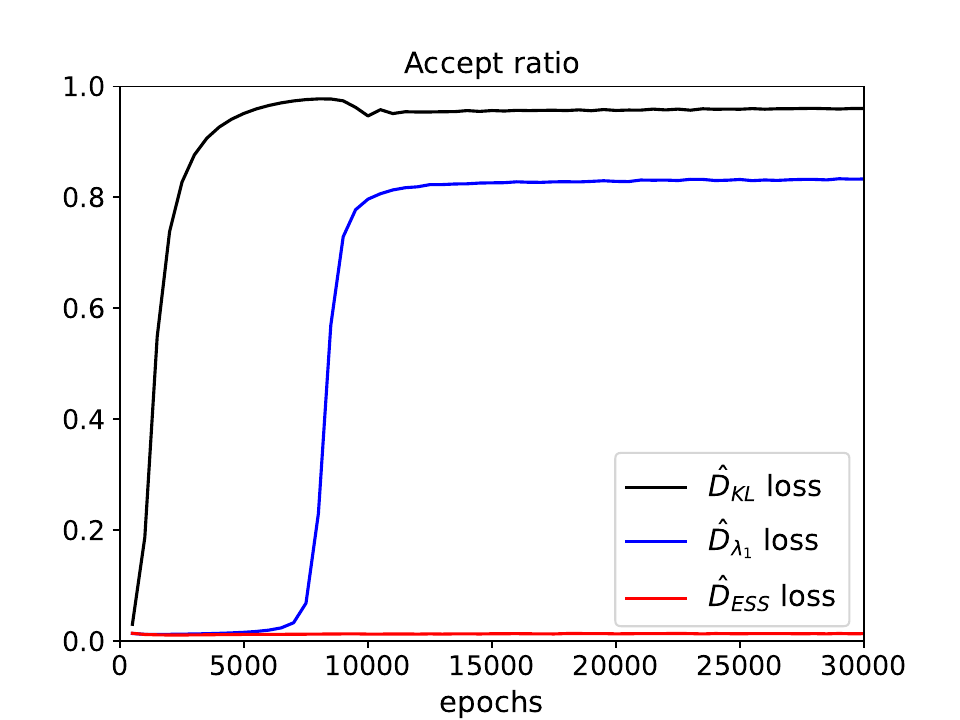}&
\includegraphics[ width=.45\textwidth]{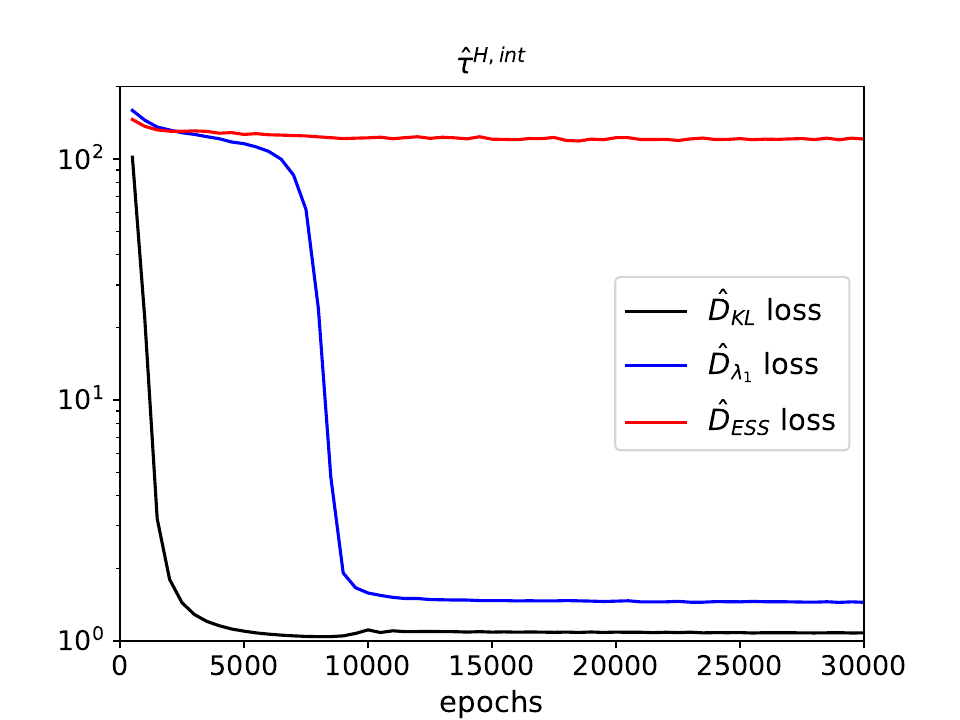}\\
\includegraphics[ width=.45\textwidth]{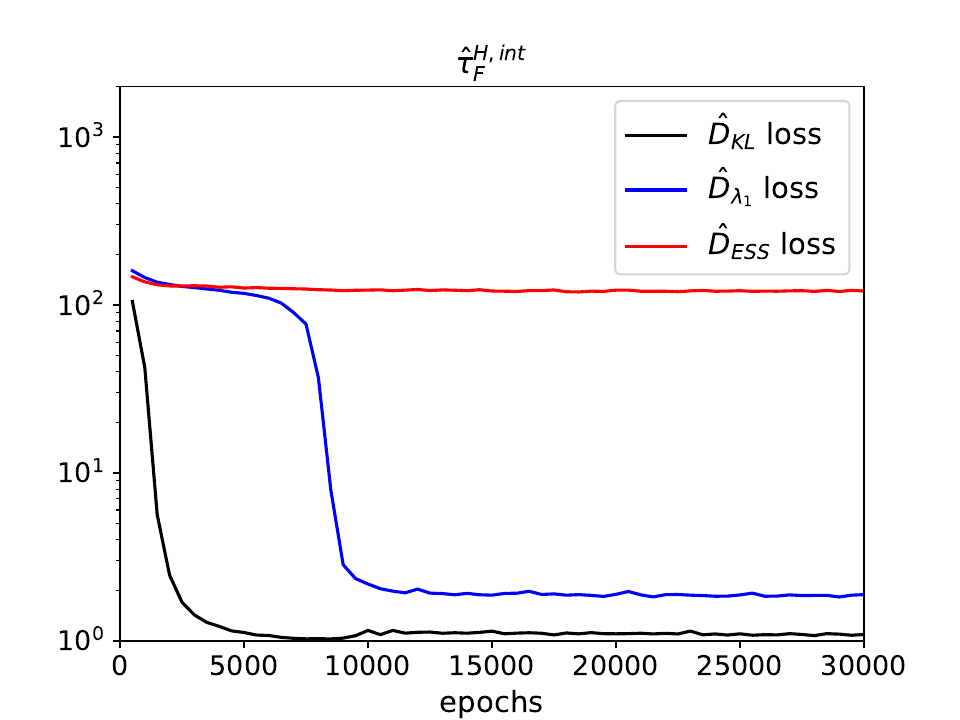}&
\includegraphics[ width=.45\textwidth]{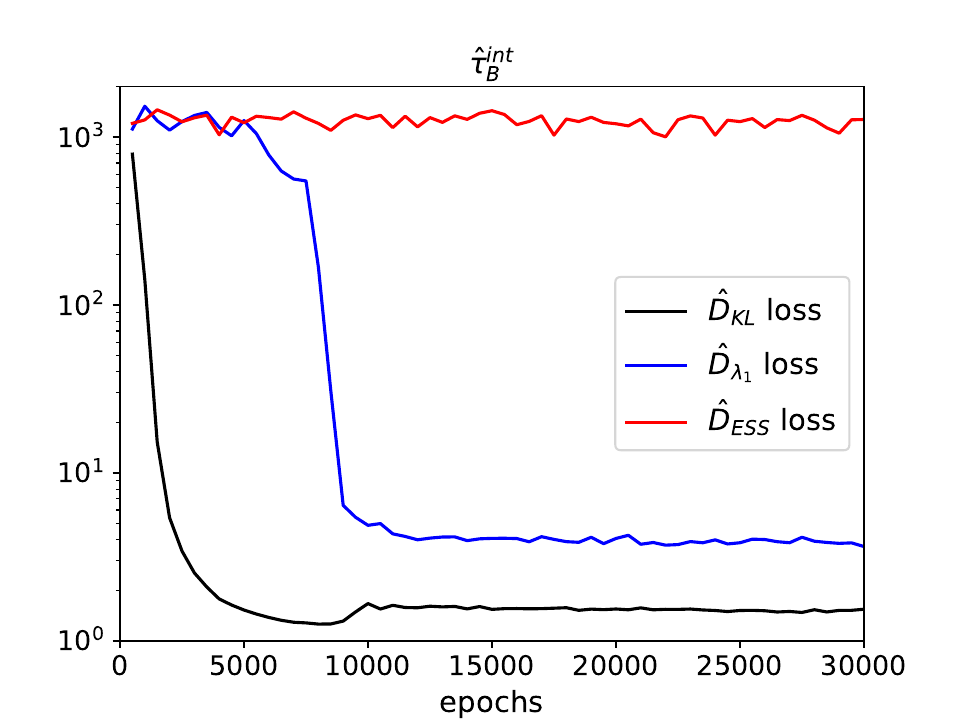}
\end{tabular}
\caption{
Learning process for $16\times 16$ system at $\beta=0.3$ using three loss functions. Different observables have been shown as functions of training epochs. Functions shown on the plots are averages of 16 learning process in order to exclude effects of random net initialization. The free energy for training with $\dess$ did not converge and is out of the scale of the figure.
}
\label{loss_funct_comparison03}
\end{figure*}

We have mentioned in Section \ref{section: loss functions} that the different loss functions are sensitive to different regions of the support of $q_{\theta}$ probability distribution. In order to try to quantify the effect of these differences on the observables accessible from the level of a simulation we now take a closer look at six quantities: the free energy $\hat F_q$, $\widehat{ESS}$, acceptance ratio in NMCMC and three definitions of the integrated autocorrelation times. We show all of them as a function of training epochs, with each of them calculated with the three loss functions introduced earlier. In order to have a better perspective we compare the situation at $\beta=0.6$ (Fig. \ref{loss_funct_comparison}) and $\beta=0.3$ (Fig.~\ref{loss_funct_comparison03}). All data points have been averaged over 16 simulations with different, random initial weights of the neural network.
We also stress that at $\beta=0.3$ we impose the translational symmetry (to be described and discussed in Section \ref{section: symmetries}). 

In the first panel of Fig.~\ref{loss_funct_comparison} we present the variational batch estimate of the free energy. The green line denotes the exact result and one clearly sees that the Kullback–Leibler divergence loss function is the most effective, i.e. provides the best approximation at a given number of training epochs. The $\dess$ loss function fails to converge to the expected value for this larger system (results for this function are very out of scale of the figure). The authors of Ref.~\cite{2021arXiv210108176A} also note in the context of normalizing flows that this function is unsuitable for training. The same behaviour of the discussed loss functions is reproduced in other observables. In other words, the better approximated is the free energy, the larger is the $\widehat{ESS}$ coefficient, the larger is the acceptance ratio and the smaller is the integrated autocorrelation time $\tI$. Also other estimates of the autocorrelation time, $\tF$ and $\tB$ exhibit the same ordering. Analogous conclusions can be drawn from Fig.~\ref{loss_funct_comparison03} where we present simulations performed at $\beta=0.3$.  Note that the $\dess$ loss functions is unable to show any improvement over training period.  

We note that in all panels of Figs.~\ref{loss_funct_comparison} and \ref{loss_funct_comparison03} the observables behave monotonously with the training time. Some of them show a scatter related to the variations between consecutive batches, however this does not affect the general trend. In particular, we do not see any period of stagnation of the training process.

When comparing the two most effective loss functions, $\dkl$ and $\dlam$, we observe that the failure to reproduce the correct value of the free energy by $\dlam$ by $0.5$ \textperthousand \ corresponds to an increase of the integrated autocorrelation time $\tI$ by a factor 2.

\subsection{Critical slowing down}

\begin{figure}
\includegraphics[ width=.48\textwidth]{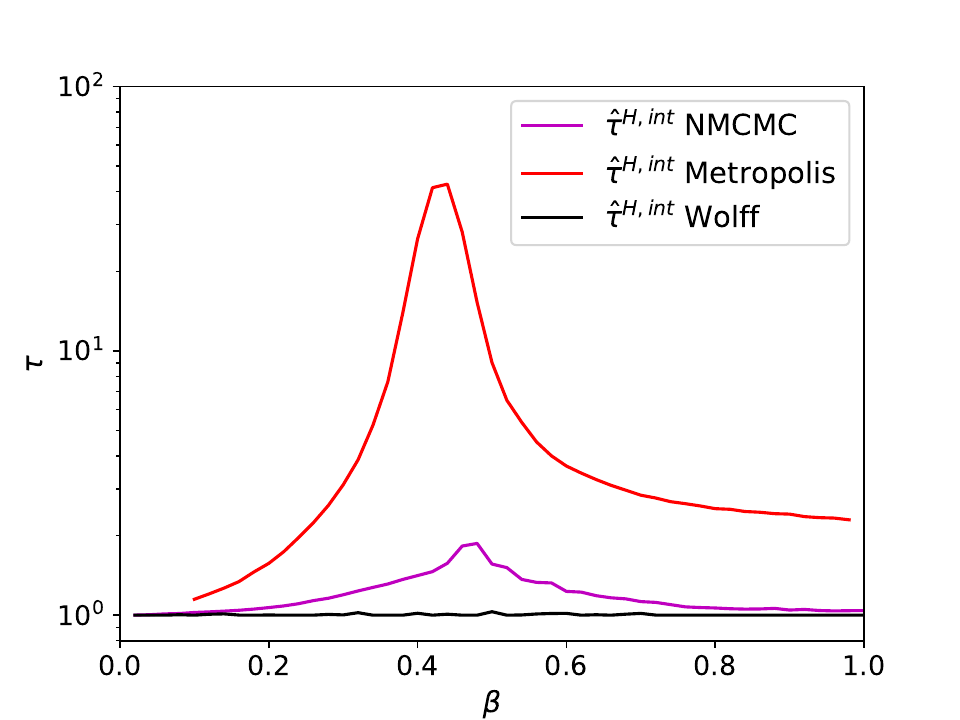}
\includegraphics[ width=.48\textwidth]{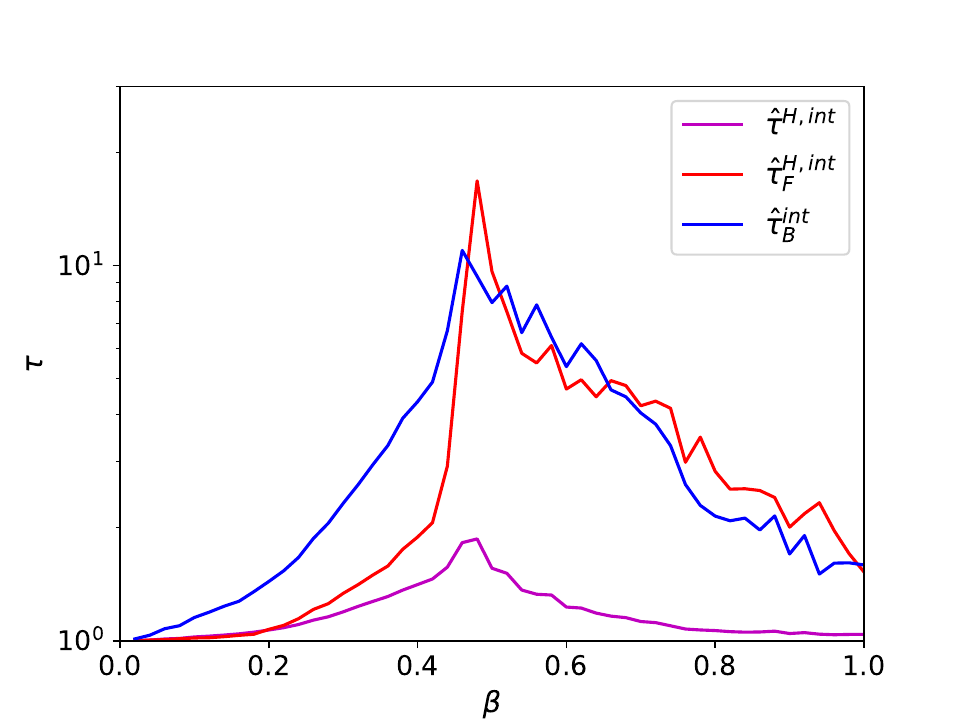}
\caption{Top: comparison of the autocorrelation time $\tI$ between different algorithms for system of size $16\times 16$ as functions of $\beta$. Bottom:
different autocorrelation times in NMCMC measured at epoch 30000, when network was fully trained. 
}
\label{taus_fig_16x16}
\end{figure}

The central issue of the MCMC simulations is the critical slowing down -- the increase of autocorrelation time when system is simulated closer to the phase transition. In the traditional approach to the simulations of the Ising model with  algorithms built upon local updates this can be very pronounced as is shown in the top panel of Fig.~\ref{taus_fig_16x16}. For the sake of comparison, we show there the integrated autocorrelation time $\tI$ measured for the Markov chain obtained using three algorithms: the NMCMC, the original Metropolis algorithm \cite{metropolis} and the Wolff's cluster algorithm \cite{PhysRevLett.62.361, WOLFF1989379}. The latter is a well-known case of cluster algorithm where the critical slowing down was completely eliminated. In order to compare the efficiencies of the Wolff and Metropolis algorithm we measured the energy after one sweep (a sweep consists of $L^2$ proposals so that each spin of the lattice is updated). In the case of cluster algorithm one sweep was defined as $L^2$ updates of clusters initiated at each spin of the lattice. 

The autocorrelation time $\tI$ in the NMCMC algorithm (magenta curve) clearly exhibits a increase around $\beta \approx 0.44$ with the maximum value close to 2. Still, such an increase is much less pronounced than the one of the Metropolis algorithm (red curve). Although, from the presented comparison the cluster algorithm (black curve) is clearly preferred, one should remember that there exist relevant physical system, for instance lattice field theories with fermionic degrees of freedom, for which a cluster formulation is unknown, whereas there are no limitations in principle in implementation of the NMCMC approach for other systems.

In the lower panel of Fig.~\ref{taus_fig_16x16} we show, for NMCMC, the dependence of various estimates of the integrated autocorrelation time on $\beta$. A similar result was discussed in the context of the small system size in Fig.~\ref{taus_fig_4x4} where no significant dependence on $\beta$ was seen for these autocorrelation times. Here we clearly observe a maximum around $\beta \approx 0.44$ for all three quantities.

\section{Reduction of autocorrelation time}
\label{sec: reduction}

In the NMCMC approach the autocorrelation times can be reduced by improving training hence reducing the difference between $q_\theta$ and $p$. Here we shall discuss two ideas which help obtaining this goal: i) making use of the symmetries of the model and ii) reducing effective number of spins using nearest neighbour interaction property. We shall test those two approaches using larger system size, i.e. $16\times 16$ spin lattices.

\subsection{Impact of symmetry on training efficiency and autocorrelation times}
\label{section: symmetries}

In the NMCMC algorithm discussed so far we did not discuss any of the symmetries of the simulated system. We consider only global transformations of configurations which do not change their energies. It immediately follows that configurations related by a symmetry have equal target probabilities. As we deal with discrete systems on a discrete and finite lattice we can limit ourselves to finite symmetry groups consisting of $I$ elements $\{\mathcal{S}_0,\mathcal{S}_1,\ldots,\mathcal{S}_{I-1}\}$, where $\mathcal{S}_0 \equiv 1$ is identity transformation $\mathcal{S}_0 \mathbf{s}= \mathbf{s}$.  The transformations $\mathcal{S}_i$ leave the energy unchanged, i.e. $H(\mathbf{s})=H(\mathcal{S}_1 \mathbf{s})=\ldots=H(\mathcal{S}_{I-1} \mathbf{s})$. Symmetries are not present in the conditional probabilities parametrized by the neural network and hence there is no obvious way of imposing them on the output of the neural network. The approach suggested in Ref.\cite{2019PhRvL.122h0602W} was to force the network to treat equivalent configurations as equally probable by symmetrizing their approximated probabilities within the batch. For that purpose, for each configuration $\mathbf{s}$ proposed by the neural network we replace its probability $q(\b s)$ by the following average:
\begin{equation}
    q_\theta(\mathbf{s})\rightarrow \frac{1}{I} \sum_{i=0}^{I-1}  q_\theta(\mathcal{S}_i \mathbf{s})
    \label{symmetrization_def}
\end{equation}
where $q_\theta(\mathcal{S}_i \mathbf{s})$ are obtained by invoking the neural network on the transformed spin configuration. This induces an additional numerical cost because for each configuration from the batch we need to call the neural network $I-1$ more times to estimate the probabilities of transformed configurations. However, in order to generate the original configuration $\b s$ we have to evaluate $L^2$ conditional probabilities, meaning that we have called the neural network $L^2$ times. Hence, if $I \ll L^2 $ the additional cost of imposing the symmetry is subleading, whereas the benefits may be significant, as we discuss below.\footnote{Note that once the configuration is generated, each symmetry transformation requires evaluation of network once. Therefore, the total number of network evaluation is $L^2+I$, where $I$ is number of symmetry transformations we apply (including identity transformation).}

In the case of the Ising model on a square lattice with periodic boundary conditions the total symmetry group $S$ is given by the outer semidirect product of the dihedral $D_4$ group of the square lattice, two cyclic permutation groups $\mathbb{Z}_L$ corresponding to translations in the $x$ and $y$ directions and the $\mathbb{Z}_2$ symmetry in the space of internal degrees of freedom of each spin, 
\begin{equation}
    S = D_4 \otimes \mathbb{Z}_L \otimes \mathbb{Z}_L \otimes \mathbb{Z}_2.
\end{equation}
The group $S$ has rank $16L^2 \equiv I$. In the following we choose two symmetries: the translational symmetry in the $y$ direction and the $Z_2$ symmetry and discuss the impact of their imposition during training through Eq.\eqref{symmetrization_def}.
\begin{figure*}
\centering
\begin{tabular}{ll}
\includegraphics[ width=.45\textwidth]{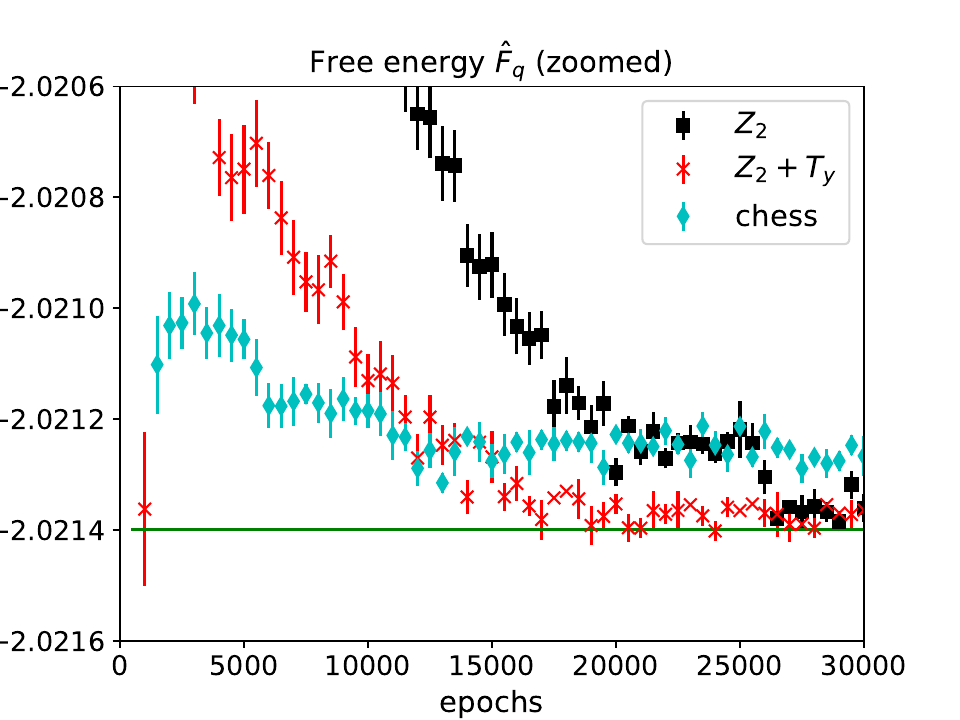} 
\includegraphics[ width=.45\textwidth]{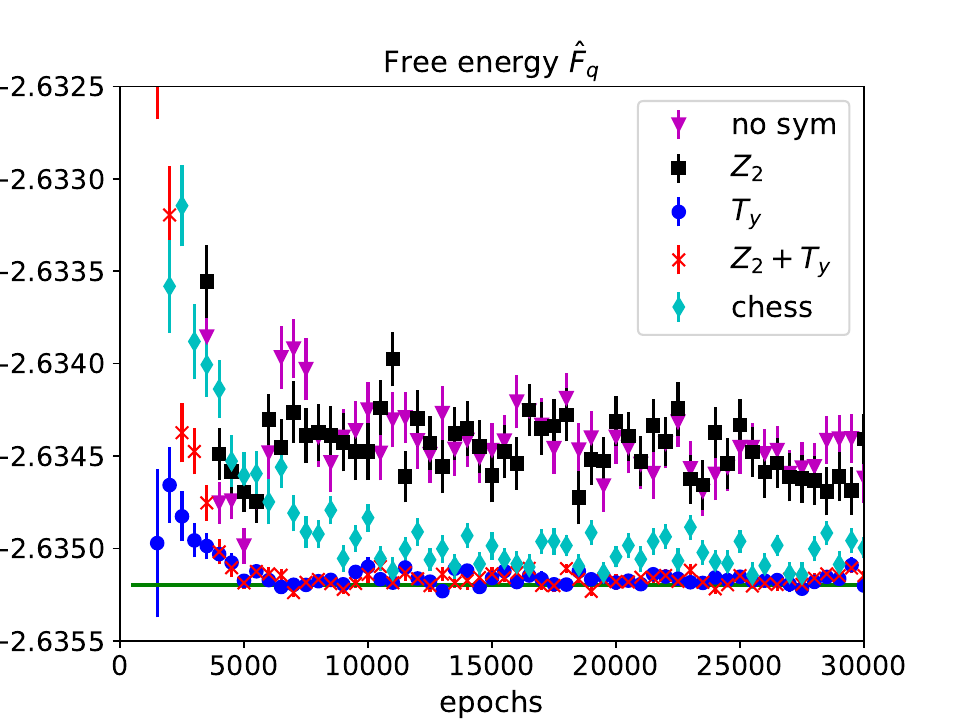} 
\end{tabular}
\caption{Free energy as function of training epochs for system $16\times 16$ and $\beta=0.6$ (left panel) and $\beta=0.3$ (right panel). For each quantity we compare the quality of the learning process for different symmetries and with chessboard factorization. On left panel the free energy for training without symmetries and with $T_y$ symmetry is out of plot range, for all points see first panel of Fig.~\ref{symm_fig_16x16_app}.
}
\label{symm_fig_16x16}
\end{figure*}
Since the symmetries ${Z_2}$ and ${T_y}$ are independent they can be imposed separately or simultaneously. We show their effects in Fig.~\ref{symm_fig_16x16} for $\beta=0.6$ and $\beta=0.3$
by discussing the free energy $\hat F_q$ (for similar plots with the acceptance ratio in NMCMC, $\widehat{ESS}$ and $\tI$ see Figs.~\ref{symm_fig_16x16_app} and \ref{symm_beta03_fig_16x16_app} in the Appendix). Each figure contains five sets of data: results from the NMCMC simulation without the imposition of any of the system symmetries, NMCMC simulation with the $Z_2$ symmetry alone, NMCMC simulation with the $T_y$ symmetry alone, NMCMC simulation with both $Z_2$ and $T_y$ symmetries imposed simultaneously and finally a data set obtained from a NMCMC simulation where the heathbath updates were incorporated on a chessboard lattice. The details of the latter are described in the next subsection. Figures showing the free energy contain additionally a green line at the exact value. Plots show the evolution of the observable as a function of the training period, so we expect convergence of the different data sets at the right edges of the figures.

The left panel of Fig.~\ref{symm_fig_16x16} shows the results obtained at $\beta=0.6$ when the system is expected to be in the ordered phase. From the first plot one readily concludes that the imposition of $Z_2$ symmetry is crucial. Data sets corresponding to simulation without this symmetry  converge very slowly or do not converge at all towards the exact value of the free energy. Only when we include the $Z_2$ symmetry the algorithm is able to reproduce the expected value (data not visible on the plot are out of the scale).

The situation is different at $\beta=0.3$ which we show in the right panel of Fig.~\ref{symm_fig_16x16}. At this temperature the system is expected to be in the disordered phase and the role of $Z_2$ and $T_y$ symmetries is reversed. In this situation the imposition of the $Z_2$ symmetry does not produce any visible improvement: for all observables the data set obtained from the simulation without any symmetry and with the $Z_2$ symmetry only are barely distinguishable. It is now the translational symmetry which improves the performance of the algorithm. Let us also note that the training is much more efficient than it was at $\beta=0.6$ as our observable reaches satisfactory values already after 10000 epochs. 

We may provide an intuitive picture behind the different roles of the symmetries in the different phases of the system. The configuration in the ordered phase has large clusters of spins pointing in the same direction, hence strongly correlated. Most of the probabilities generated by the neural network have to be close either to 0 or to 1. In that case the $Z_2$ symmetry changes the conditional probability $\epsilon \rightarrow 1-\epsilon$ or $1-\epsilon \rightarrow \epsilon$, which is highly nontrivial from the neural network point of view. On the contrary, the translational symmetry shifts the spins in some direction. However, since most of them are pointing in the same direction, this may result in a new configuration which is essentially equivalent to the original one. Therefore, the imposition of the translational symmetry has a small impact on the training quality.

As far as the disordered phase is concerned, the situation is reversed. In this phase most spins are random and uncorrelated between each other. Hence, most of conditional probabilities produced by the neural network should be around $\frac{1}{2}$. In this particular situation, the $Z_2$ symmetry leaves the configuration and the conditional probabilities essentially unchanged as $\frac{1}{2} \rightarrow 1 - \frac{1}{2} = \frac{1}{2}$. Hence, one would expect the effect of the imposition of the $Z_2$ symmetry to be small. The translational symmetry changes the configuration by shifting the spins, but as they are mostly random, one naively would expect that its effect would be also small. However, non-trivially we found that from the neural network point of view such a transformation carries some important information, and the imposition of the symmetry improves the training significantly.

Comparing the results shown in Fig.~\ref{symm_fig_16x16}  one concludes that the imposition of system symmetries is pivotal in order to make the training process more efficient. Additionally, we observe that the different symmetries play different role in the various regions of the phase space of the system, and hence one should implement as many of them as possible. Let us highlight again that in the case of two-dimensional systems, as the system size grows, the number of invocations of the neural network scales like $L^2$ as one needs to evaluate the conditional probability of each spin. On the contrary, the imposition of symmetries is equivalent to a single neural network call to evaluate the total probability of the symmetry-transformed configuration and hence the cost of the symmetries depends on the number of symmetry transformations that are used to average the approximated probabilities. The latter may or may not depend on $L$.

\subsection{Chessboard factorization}
\label{section: chessboard}

Fig.~\ref{symm_fig_16x16} contained another data set which was not commented in the previous section and which we explain below. The chessboard factorization which was used to generated this data can be applied to models with only nearest-neighbours interactions. One has to divide all spins into two species, odd and even, according to the parity of the sum of their coordinates on the lattice, this results in a "chessboard" pattern. From the locality of the interaction it turns out that the total probability of that spin from an even site depends exclusively on the state of the four neighbouring spins from the odd sites. In other words, if all spins at odd sites have been fixed, the spins at even sites can be generated from the exact, local Boltzmann probability distribution. 

In the present context the above formulation means that  one can reduce the number of spins which have to be set by the neural network by a half. Not only this reduces significantly the number of invocations of the neural network, but also the size of the neural network is reduced and more importantly half of the spins can be generated independently from each other from the target probability distribution further decorrelating the system. 

The formula Eq.~\eqref{eq:conditional_probabilities} is now modified:
\begin{align}
    q_\theta(\mathbf{s}) = \prod_{i=1}^{N/2} q_O(s^i| s^{i-1} \dots s^1) \prod_{i=N/2+1}^{N} q_E(s^i, n(s^i) ) ,
    \label{conditional_probabilities_chess}
\end{align}
where $q_O(s^i| s^{i-1} \dots s^1)$ is a probability of $i$-th \emph{odd} spin to be $s^i$ when other \emph{odd} spins $s^{i-1} \dots s^1$ are fixed and $q_E(s^i, n_A(s^i)$ is probability of $i$-th \emph{even} spin to be $s^i$ when it's neighbours $n(s_i)$ are fixed (note that even spins have only odd neighbours). The procedure to get the full spin state is the following: i) run forward network to get odd spins, ii) using values of fixed odd spins to calculate probabilities $q_E$ of all even spins, iii) draw the even spins using calculated $q_E$. 
Probability of $i$-th even spin to be $+1$ or $-1$ is equal:
\begin{align}
q_E(s^i=\pm 1, n(s^i))=  \left[ 1+\exp \left(\mp\, 2\beta  \sum_{\ j \in n(s^i) } s^j \right) \right]^{-1},
\end{align}
where the sum is performed over nearest neighbours $n(s^i)$ of spin $s_i$.

As the neural network is used to generate only half of spins in full state we expect that training is faster and more effective when chessboard factorization is used. In Fig.~\ref{symm_fig_16x16} we show (light blue diamonds) observables during the training when this factorization is applied, no symmetries involved here (so they should be compared with magenta triangles). We observe dramatic difference in performance of the network - chessboard factorization improves significantly all measures of training quality, comparable to that of the symmetries imposition.

\section{Summary and outlook}
\label{section: summary}

In this work we have investigated, from many perspectives, the neural Markov Chain Monte Carlo algorithm for simulating statistical systems with a finite number of discrete degrees of freedom. Although we have concentrated on the two-dimensional Ising model on a square lattice, most of our observations carry over to simulations of more complex systems, for example spin glass models. We quantified the performance of the new algorithm by measuring various estimators of integrated autocorrelation times and discussed their behaviour as a function of inverse temperature $\beta$ and neural network training time. In particular, we have verified that the NMCMC algorithm is mildly affected by the critical slowing down close to the phase transition, i.e. the integrated autocorrelation time does not increase considerably in this region. Of course, this is only true if the neural network was trained well enough so that the encoded probability distribution is sufficiently close to the target one. We have highlighted the link between the NMCMC algorithm and the Metropolized Independent Samplers studied by Liu where exact results for the leading autocorrelation time exist in terms of importance ratios. By exploiting this correspondence we suggested some measures of the quality of the training of the neural network based on importance ratios. In particular, the width of the histograms of importance ratios gathered at each stage of the training procedure may help to decide when the training has been successful. 
Comparing with the local Metropolis algorithm we stress that the critical slowing down of the Metropolis algorithm can be explained by the growing size of spin clusters in the parameter region around the phase transition, hence has a physical interpretation. In the case of the NMCMC algorithm the non-zero autocorrelation times are due  to an imperfect approximation of the Boltzmann distribution by $q_{\theta}$. Therefore the phenomenon of the critical slowing down, if any,  stems from the dynamics of the neural network training. Furthermore  the configurations with very small $p$ are the ones for which the network will have biggest difficulty for getting $q_\theta$ right. This can lead to very large relative errors and hence very large values of $\tL$. As those configurations have $q_\theta(\b s)\ll p(\b s)$ they will be proposed very rarely if ever. However if they do appear they will repeated approximately $w(\b s)$ times. At this moment it is unclear to us if this may have an impact on the results of simulations, as we have not encountered such situation in practice.

As mentioned, importance ratios are the building blocks of the analytic expressions for the eigenvalues of the Markov chain transition matrix and hence are used to describe the autocorrelations in the Metropolized Independent Sampler class of algorithms. We have constructed a loss function directly proportional to the largest eigenvalue of the transition matrix of the Markov chain. Although its performance is inferior to the previously studied $\dkl$ loss function, it provides an interesting example of a loss function which behaves differently when the neural network training is implemented with the REINFORCE algorithm or with the gradient learning. This is in contrast to $\dkl$ loss function that is the same in both approaches. Hence, in further studies $\dlam$ may be used to investigate different approaches to neural network training. As a third alternative, we checked the $\dess$ loss function discussed in Ref.~\cite{2021arXiv210108176A}, and confirmed that its efficiency is very poor in the studied context.

We have investigated the impact of symmetries on the training process. It was already noted in Ref.~\cite{2019PhRvL.122h0602W} that the $Z_2$ symmetry is crucial in order to correctly reproduce the free energy of the system. In our detailed study we observed that this is true only in a specific range of inverse temperatures $\beta$. For instance, for small $\beta$ the $Z_2$ symmetry does not improve the quality of training. Instead, another symmetry turns out to be more important, namely the translational symmetry, which in turn was not as efficient at large values of $\beta$ as the $Z_2$ symmetry. Hence, we conclude that in order to efficiently train the neural network in the full range of $\beta$ it may be advisable to impose both, or all accessible, symmetries simultaneously. 

Another observation concerns the chessboard factorization, which turns out to considerably reduce both the width of the neural network and the training time. We find that its effects are as important as the effects of the symmetries and therefore it may be advantageous to incorporate it in the algorithm in the full range of $\beta$.

In summary we note that the improvements proposed and discussed in this work also apply to the variational estimation of free energy, as proposed in Ref.~\cite{2019PhRvL.122h0602W}.  In other words, the described hints which allow to better train the neural network, have as immediate consequence that the systematic deviation of the free energy from its exact value is much smaller.

We finish by noting that the main concern towards the practical application of the above method may be its scalability to larger systems. We have addressed this issue in the meantime and proposed a hierarchical approach which allows to simulate systems of size $128 \times 128$ and larger \cite{Bialas:2022qbs}. The advances presented there rely heavily on the experiences gathered and described in the present work.

 \section*{Acknowledgement}
 Computer time allocation 'plgnnformontecarlo' on the Prometheus supercomputer hosted by AGH Cyfronet in Krak\'{o}w, Poland was used through the polish PLGRID consortium. This research was funded in part by the Polish National Science Center (NCN) grants No.\,2019/32/C/ST2/00202 and 2021/43/D/ST2/03375. For the purpose of Open Access, the author has applied a CC-BY public copyright licence to any
Author Accepted Manuscript (AAM) version arising from this submission. T.S. kindly acknowledges support of the Faculty of Physics, Astronomy and Applied Computer Science, Jagiellonian University Grant No.\,2021-N17/MNS/000062. This research was funded by the Priority Research Area Digiworld under the program Excellence Initiative – Research University at the Jagiellonian University in Kraków.

\appendix

\section{Some relations among the different definitions}
\label{app. taus}

Let us comment on the hierarchy of (integrated) autocorrelation times we have defined in section \ref{section_est_at}, $\tOi$, $\tLOi$ and $\tLi$. By construction the largest of them will be $\tLi$. As mentioned earlier, it is the property of the Markov chain not associated to any particular observable. Hence, $\tLOi$ and $\tOi$ are smaller than or equal to $\tLi$. In principle $\tOi$ is equal to $\tLi$ only if the observable $\mathcal{O}$ couples exclusively to the slowest mode of the Markov chain. One can show that once the observable has a non-zero coupling to another mode, $\tOi$ will be strictly smaller than $\tLi$. 
Also by definition, $\tOi$, being the integral of the autocorrelation function, is smaller than $\tLOi$, except for the case where the observable couples exactly to one of the modes of the Markov chain. In that situation we expect that $\tOi = \tLOi$.

As for the $\htOi$, $\htOFi$, $\htBi$ estimators their hierarchy cannot be theoretically established as they are all calculated in a very different way. However we could expect that the ordering presented above is, at least approximately, preserved leading to $\htBi\ge\htOFi\ge\htOi$. We present the measurements of the above integrated autocorrelation times on a $4 \times 4$ and $16 \times 16$ lattice for the Ising model in the following sections and find that while $\htOFi\ge\htOi$, $\htBi$ can be smaller than $\htOFi$.  We would also expect that all autocorrelation times are smaller than $\tLi$ and this is indeed the case. It is a practical consequence, that one can rely on the approximated 'batch' versions of the estimates of the autocorrelation times during the training process as they allow to quantify the amount of training needed to achieve a desired integrated autocorrelation time.

\section{Dynamics of neural network training}
\label{app. dynamics}

\begin{figure*}
\centering
\begin{tabular}{llll}
{\large $\beta=0.6$:}\\
\includegraphics[ width=.22\textwidth]{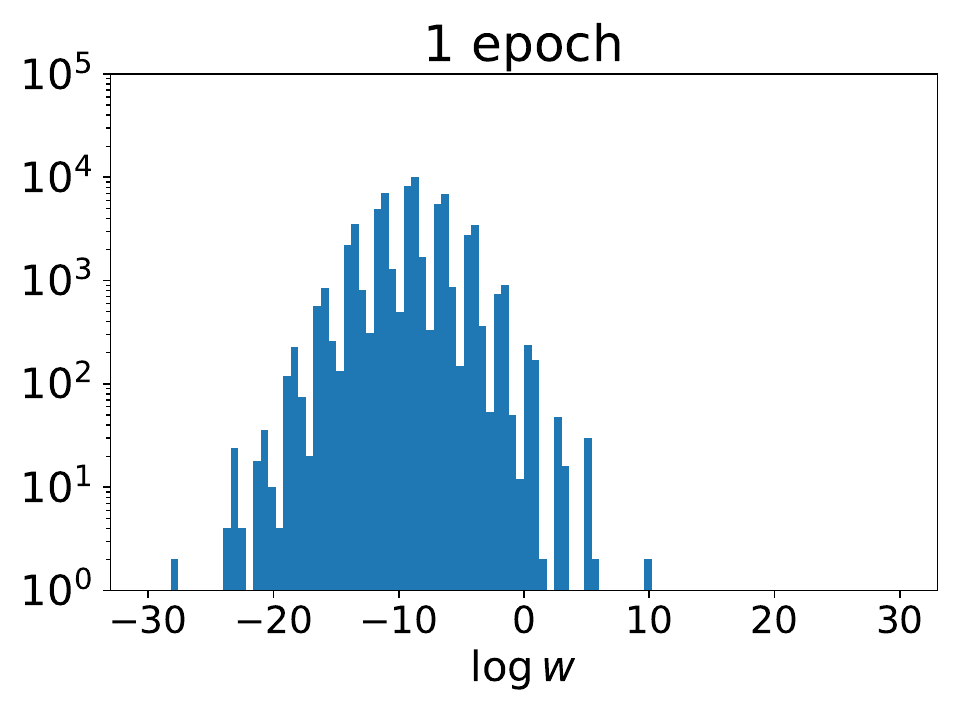} &
\includegraphics[ width=.22\textwidth]{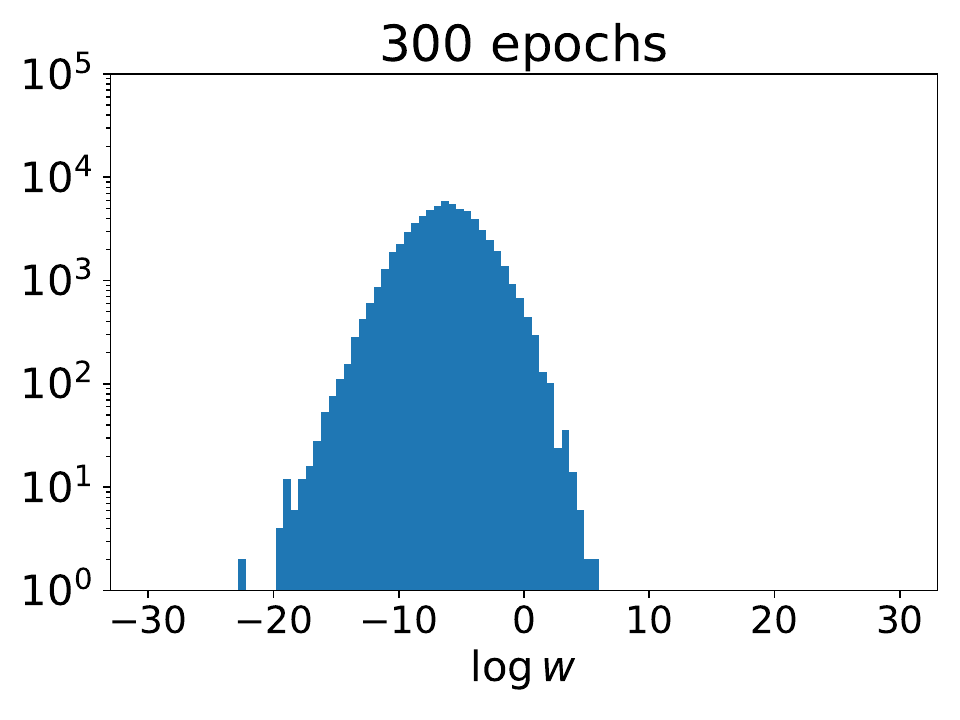} &
\includegraphics[ width=.22\textwidth]{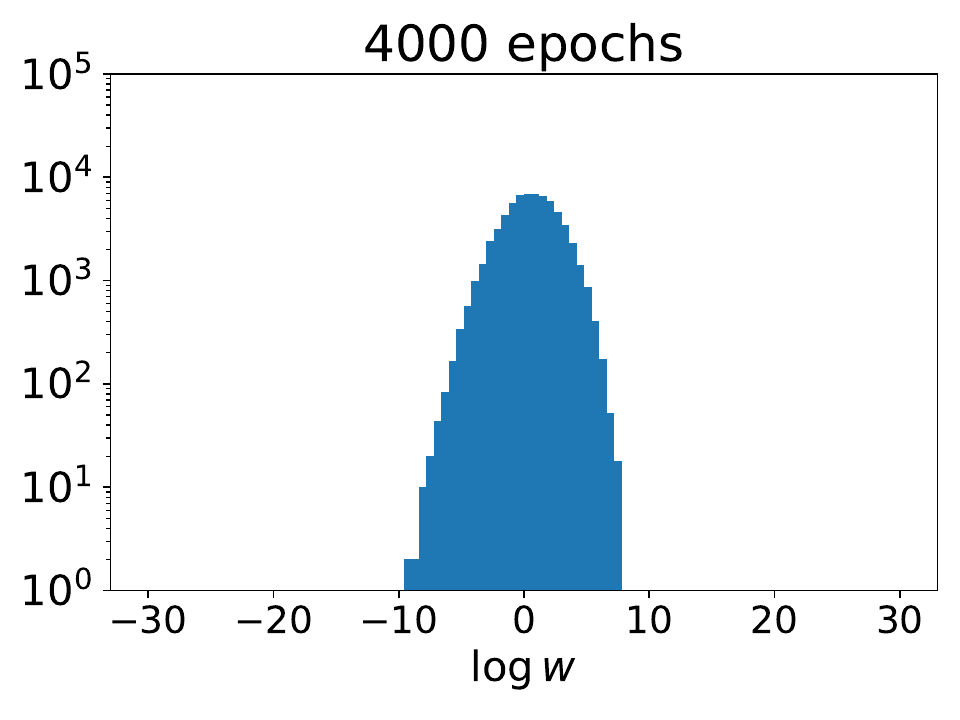} &
\includegraphics[ width=.22\textwidth]{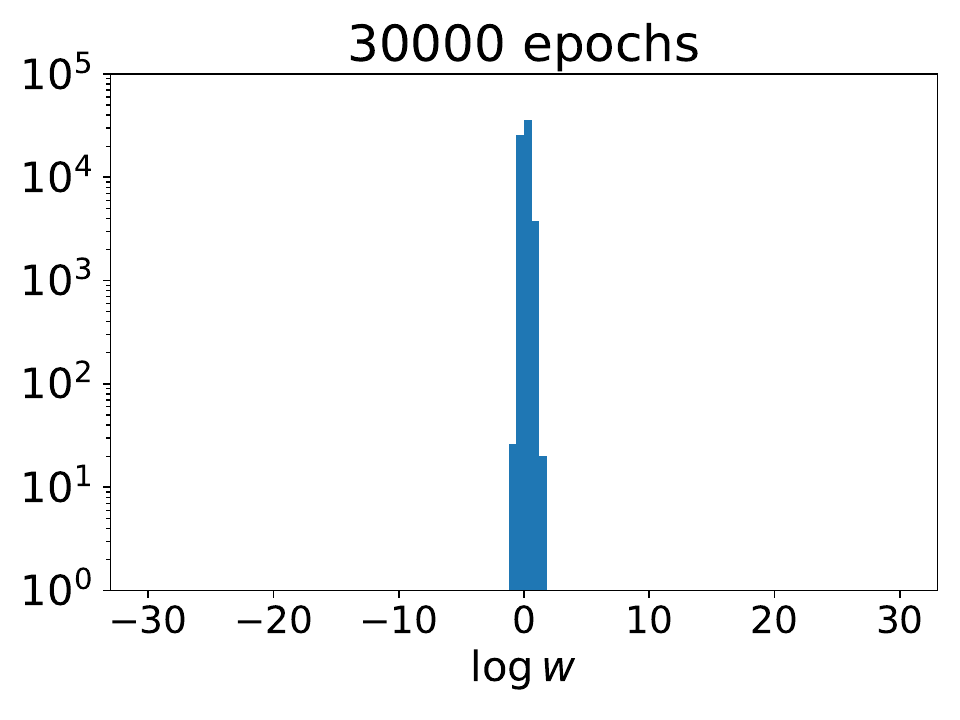}\\
{\large $\beta=0.3$:}\\
\includegraphics[ width=.22\textwidth]{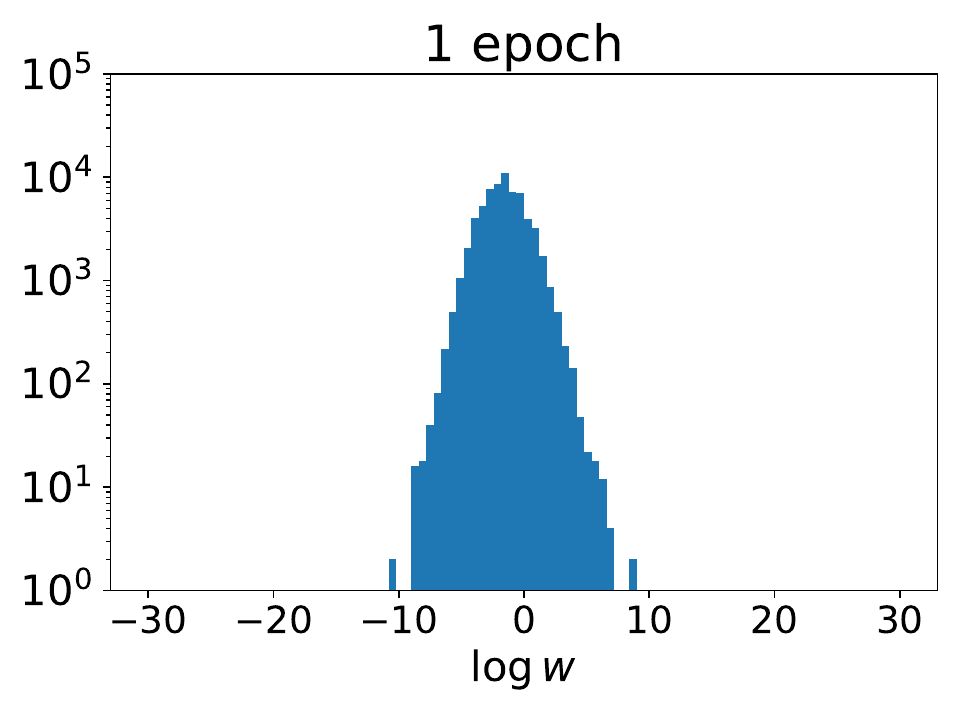} &
\includegraphics[ width=.22\textwidth]{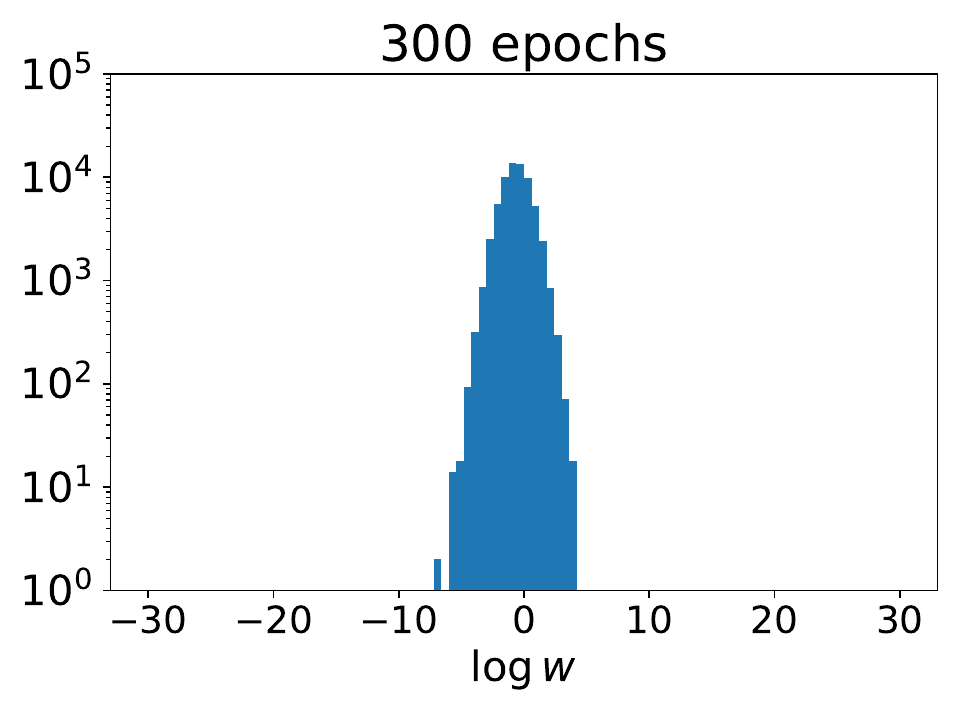} &
\includegraphics[ width=.22\textwidth]{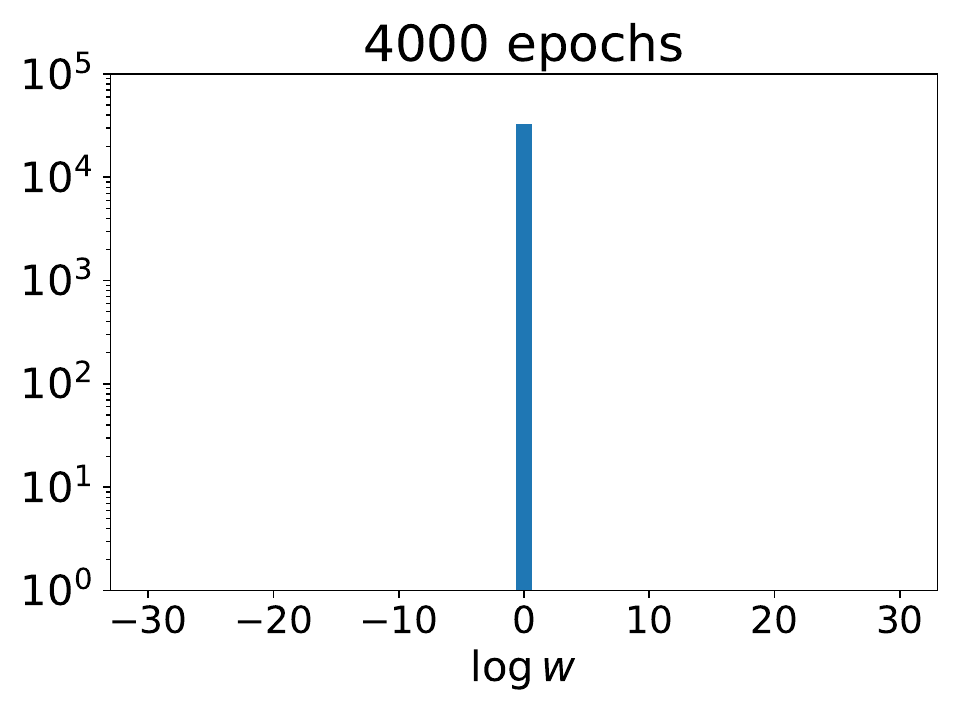} &
\includegraphics[ width=.22\textwidth]{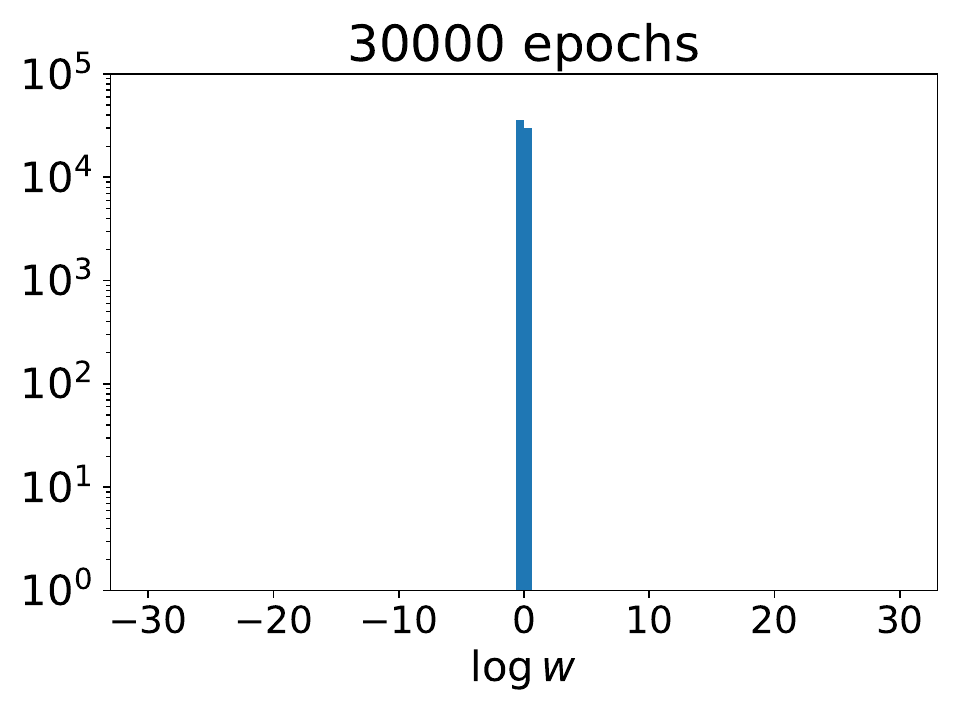}
\end{tabular}
\caption{Dynamics of the neural network training using the KL divergence loss function Eq.~\eqref{eq:DKL_loss}. For two inverse temperatures: $\beta=0.6$ (upper row) and $\beta=0.3$ (lower row) we provide snapshots after 1, 300, 4000 and 30000 epochs of histograms of $\log w_i$.}
\label{fig:dynamics logw}
\end{figure*}

The same information as plotted in Fig.~\ref{fig:dynamics logqlogp} but plotted in somewhat different manner is provided in Fig.~\ref{fig:dynamics logw} where we show snapshots of the histograms of the log of importance ratios $w$. In the ideal case, the neural network would provide $\log q_\theta $ which are equal to $\log p$ and all importance ratios would be equal to 1. Indeed, as the training proceeds the histogram is more and more peaked at $\log 1 = 0$. It is also evident that it is much easier for the neural network to learn the probability distribution $p$ at $\beta = 0.3$ than at $\beta=0.6$.

\section{Imposition of symmetries during training}
\label{app. symmetries}

This appendix contains plots similar to Fig.~\ref{symm_fig_16x16} with the convergence of the acceptance ratio in NMCMC, $\widehat{ESS}$ and $\tI$ as a function of training epochs. Each figure contains five sets of data: without and with the imposition of different symmetries and with the chessboard factorization. For the detailed discussion see the main text.

\begin{figure*}
\centering
\begin{tabular}{ll}
\includegraphics[ width=.45\textwidth]{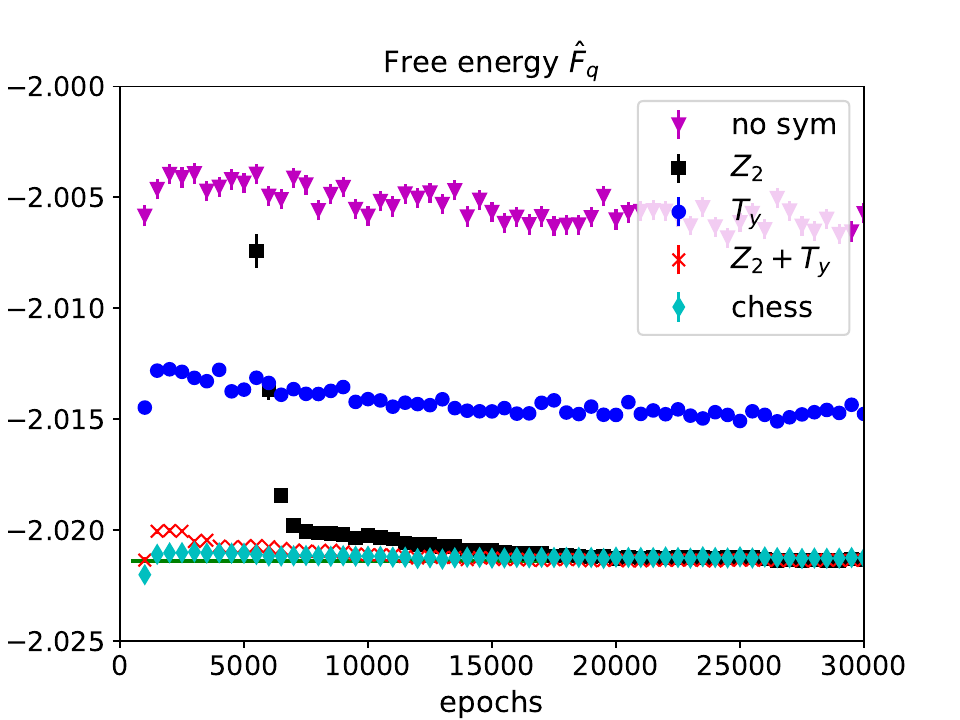}\\ %\includegraphics[ width=.45\textwidth]{F_16_symmetries_zoomed.pdf} \\
\includegraphics[ width=.45\textwidth]{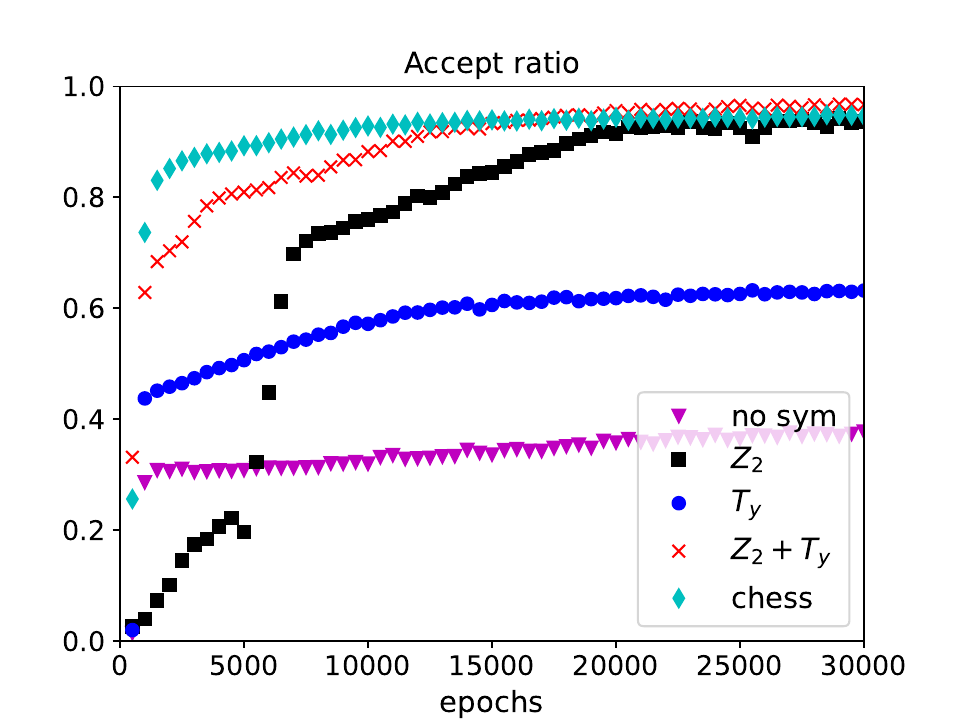} &
\includegraphics[ width=.45\textwidth]{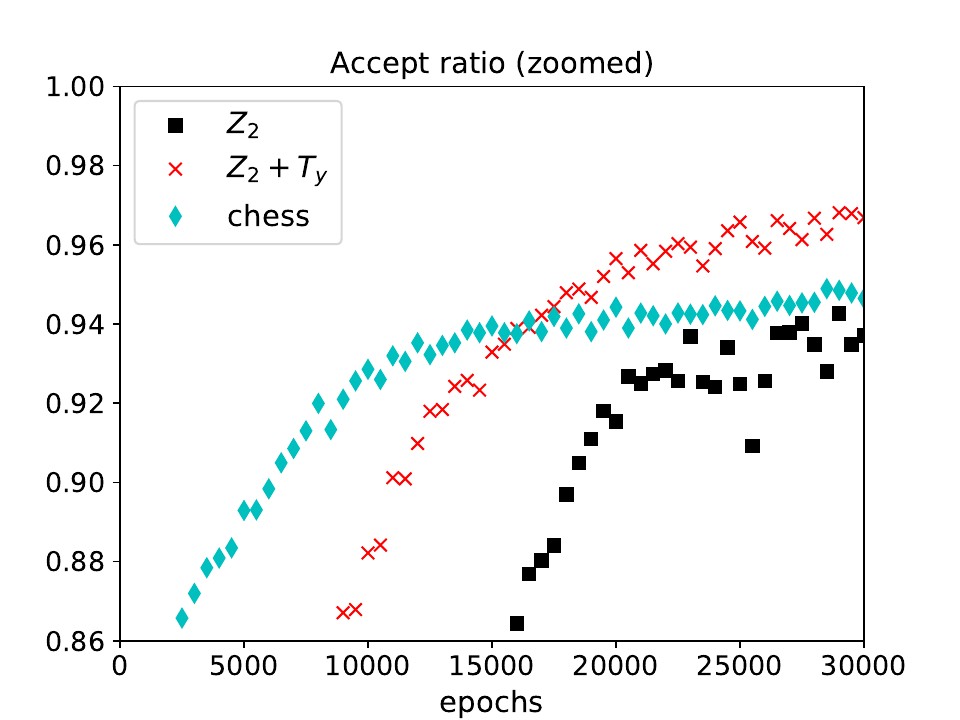} \\
\includegraphics[ width=.45\textwidth]{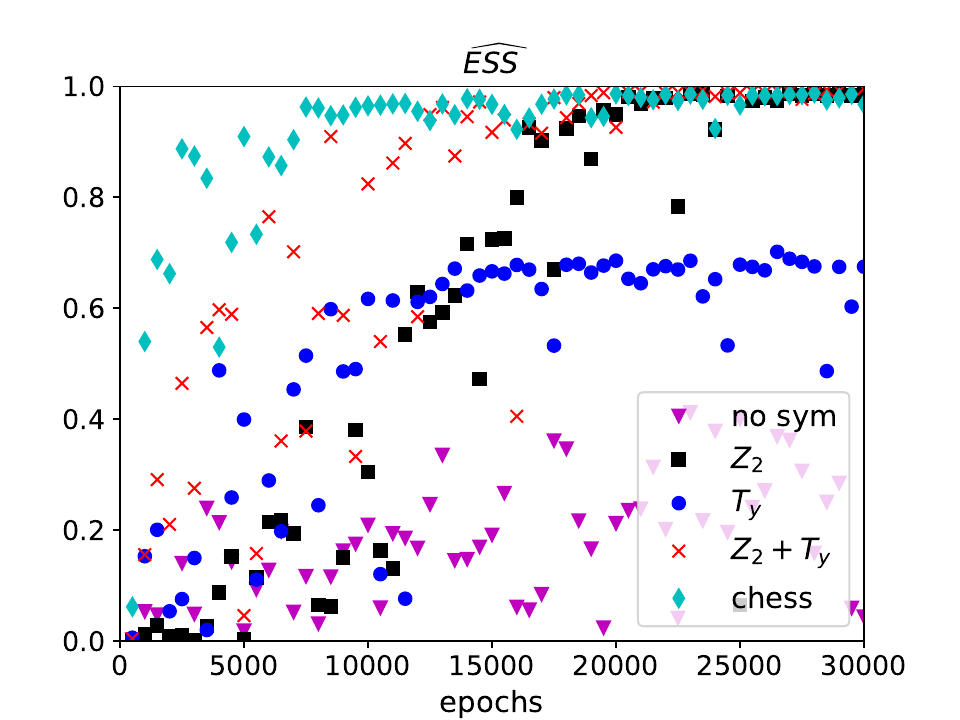} &
\includegraphics[ width=.45\textwidth]{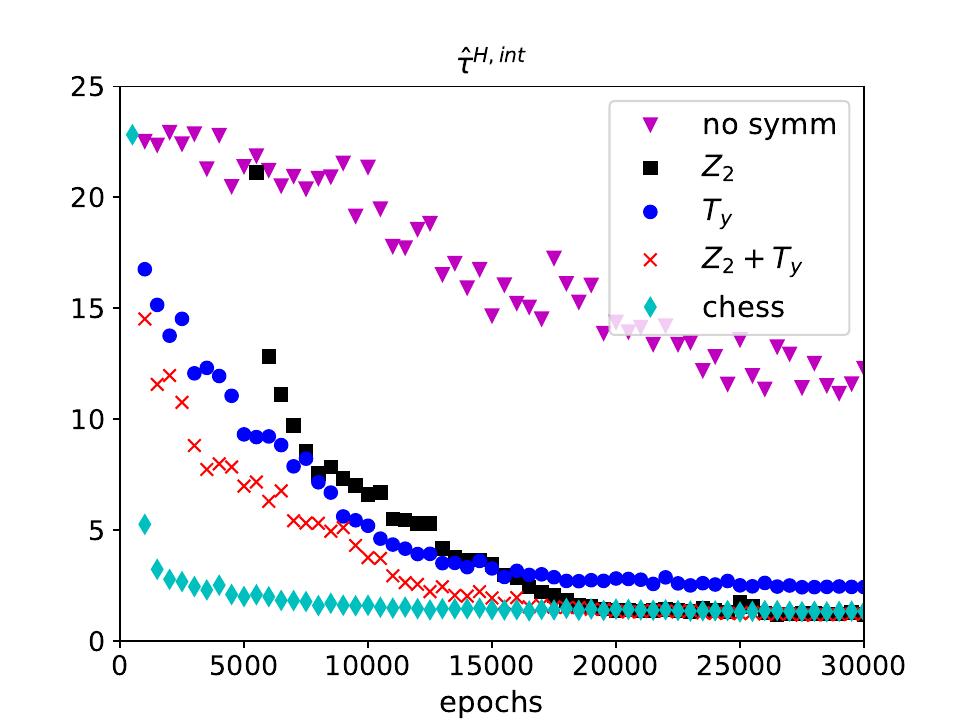}
\end{tabular}
\caption{Free energy $F$, accept ratio in NMCMC, $\widehat{ESS}$ and integrated autocorrelation time $\tI$ as functions of epochs during training for system $16\times 16$ and $\beta=0.6$. For each quantity we compare the quality of the learning process for different symmetries and with chessboard factorization. Accept ratio plot is zoomed in order to distinguish between the most efficient approaches. For zoomed plot of $F$ see Fig.~\ref{symm_fig_16x16}.
}
\label{symm_fig_16x16_app}
\end{figure*}
\begin{figure*}
\centering
\begin{tabular}{ll}
\includegraphics[ width=.45\textwidth]{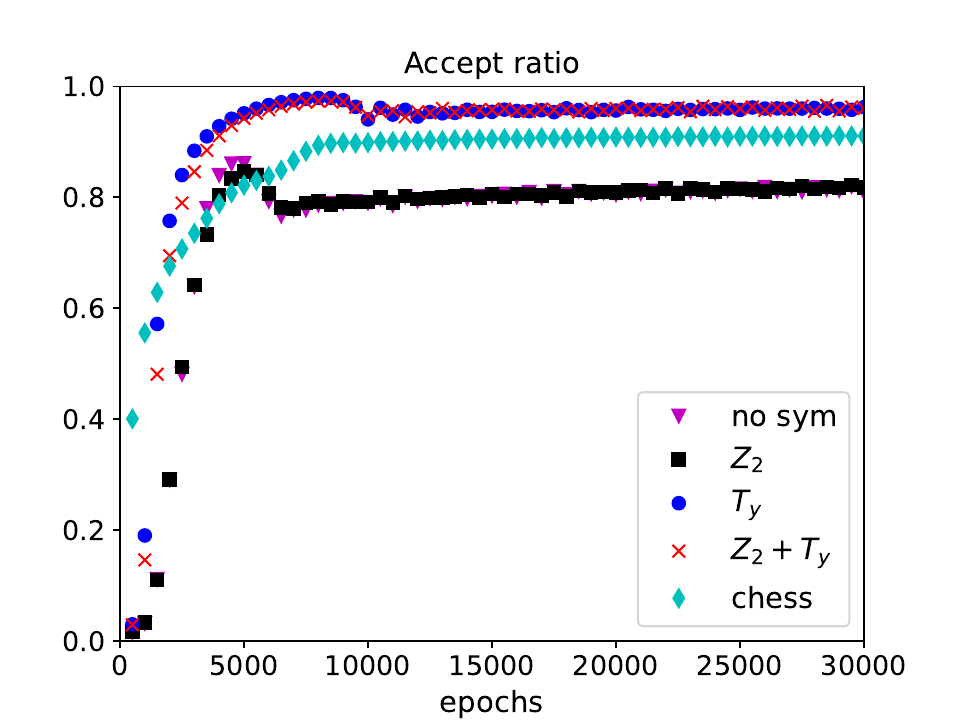} &
\includegraphics[ width=.45\textwidth]{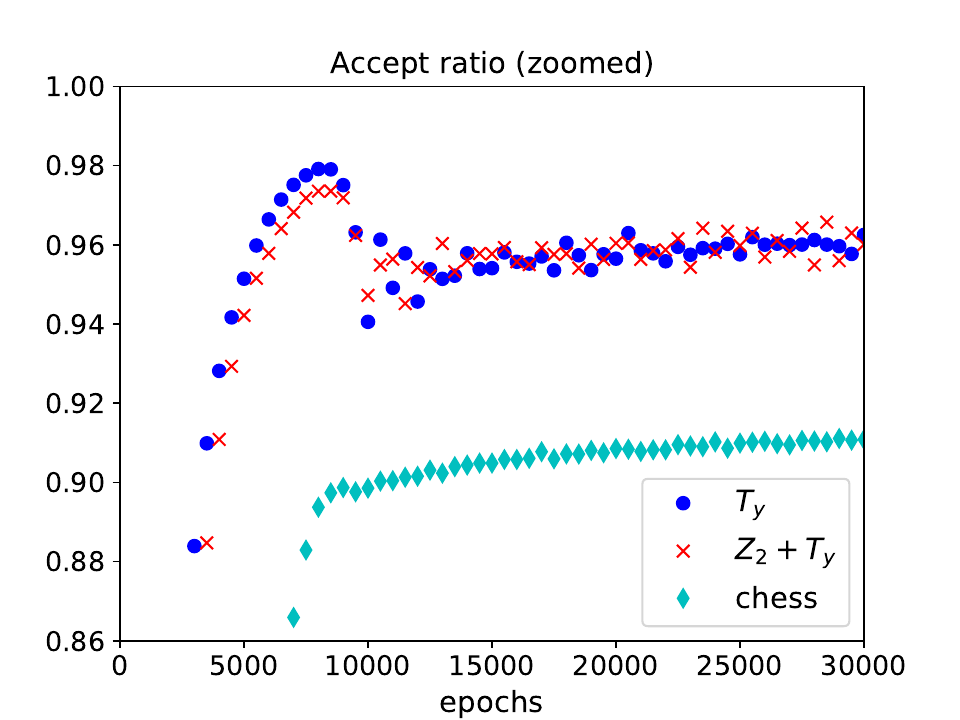} \\
\includegraphics[ width=.45\textwidth]{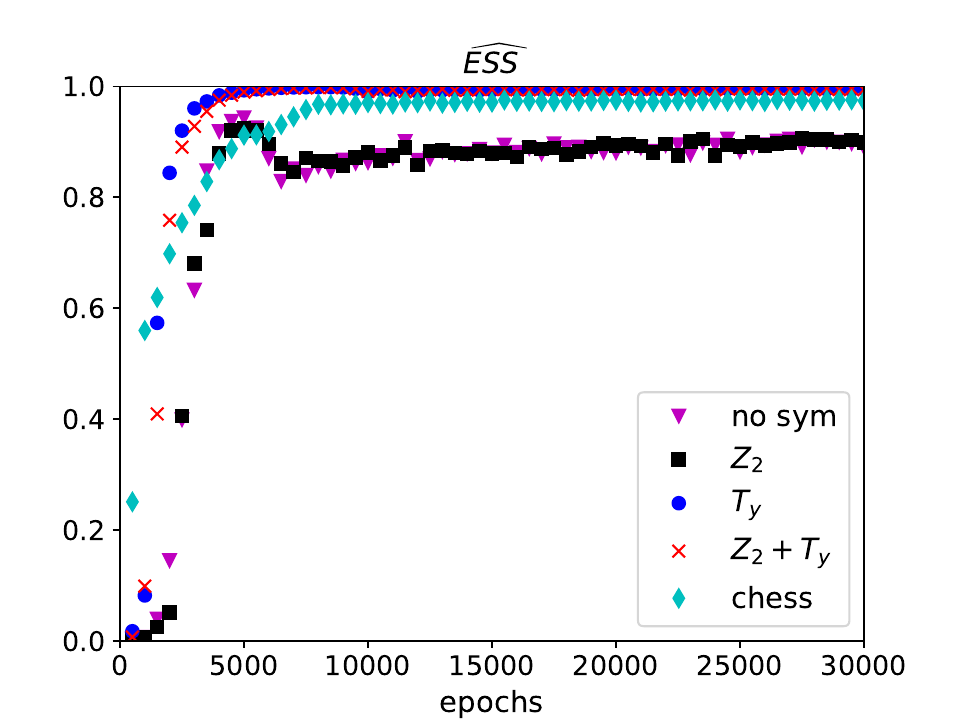} & 
\includegraphics[ width=.45\textwidth]{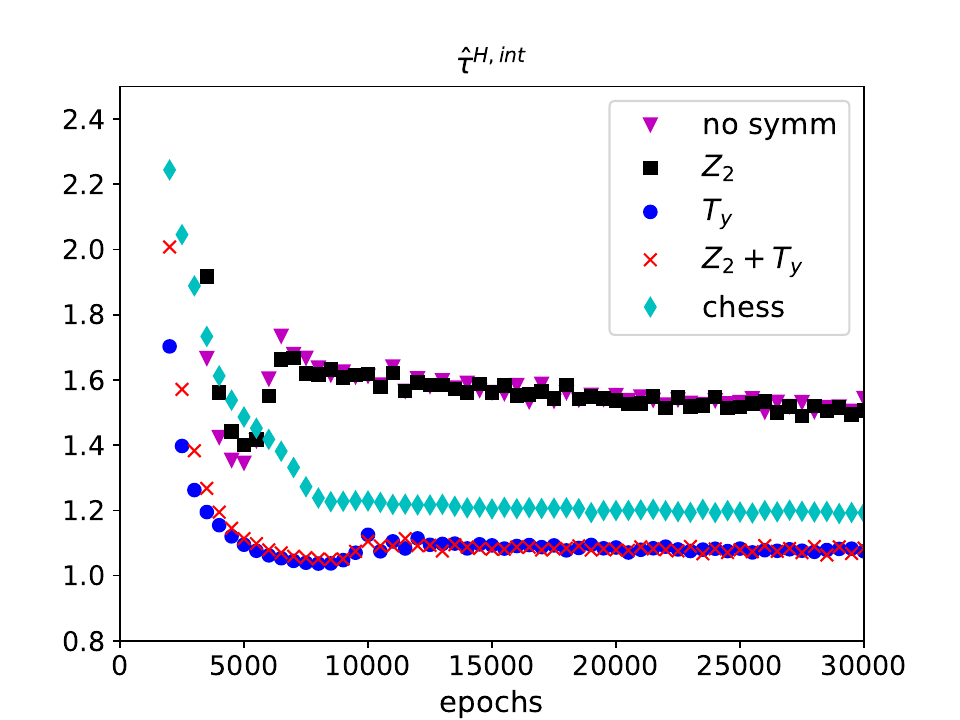}
\end{tabular}
\caption{Accept ratio in NMCMC, $\widehat{ESS}$ and integrated autocorrelation time $\tI$ as functions of epochs during training for system $16\times 16$ and $\beta=0.3$. For each quantity we compare the quality of the learning process for different symmetries and with chessboard factorization. Accept ratio plot is zoomed in order to distinguish between the most efficient approaches.}
\label{symm_beta03_fig_16x16_app}
\end{figure*}

\bibliographystyle{ieeetr}
\bibliography{references_NN}% Produces the bibliography via BibTeX.

\begin{thebibliography}{10}

\bibitem{metropolis}
N.~Metropolis, A.~W. Rosenbluth, M.~N. Rosenbluth, A.~H. Teller, and E.~Teller,
  ``Equation of state calculations by fast computing machines,'' {\em The
  Journal of Chemical Physics}, vol.~21, no.~6, pp.~1087--1092, 1953.

\bibitem{binder}
K.~Binder and D.~Heermann, {\em {Monte Carlo Simulation in Statistical Physics:
  An Introduction}}.
\newblock Springer, 2019.

\bibitem{metadynamics}
A.~{Laio} and F.~L. {Gervasio}, ``{Metadynamics: a method to simulate rare
  events and reconstruct the free energy in biophysics, chemistry and material
  science},'' {\em Reports on Progress in Physics}, vol.~71, p.~126601, Dec.
  2008.

\bibitem{metadynamics_nature}
G.~Bussi and A.~Laio, ``{Using metadynamics to explore complex free-energy
  landscapes},'' {\em Nature Reviews Physics}, vol.~2, p.~200, 2020.

\bibitem{Fucito:1984gw}
F.~Fucito, C.~Rebbi, and S.~Solomon, ``{Finite Temperature {QCD} in the
  Presence of Dynamical Quarks},'' {\em Nucl. Phys. B}, vol.~248, pp.~615--628,
  1984.

\bibitem{PhysRevD.92.114516}
M.~G. Endres, R.~C. Brower, W.~Detmold, K.~Orginos, and A.~V. Pochinsky,
  ``Multiscale monte carlo equilibration: Pure yang-mills theory,'' {\em Phys.
  Rev. D}, vol.~92, p.~114516, Dec 2015.

\bibitem{2019PhRvL.122h0602W}
D.~{Wu}, L.~{Wang}, and P.~{Zhang}, ``{Solving Statistical Mechanics Using
  Variational Autoregressive Networks},'' {\em Phys. Rev. Lett.}, vol.~122,
  p.~080602, Mar. 2019.

\bibitem{2020PhRvE.101b3304N}
K.~A. {Nicoli}, S.~{Nakajima}, N.~{Strodthoff}, W.~{Samek}, K.-R. {M{\"u}ller},
  and P.~{Kessel}, ``{Asymptotically unbiased estimation of physical
  observables with neural samplers},'' {\em Phys. Rev. E}, vol.~101, p.~023304,
  Feb. 2020.

\bibitem{PhysRevD.100.034515}
M.~S. Albergo, G.~Kanwar, and P.~E. Shanahan, ``Flow-based generative models
  for markov chain monte carlo in lattice field theory,'' {\em Phys. Rev. D},
  vol.~100, p.~034515, Aug 2019.

\bibitem{2021arXiv210108176A}
M.~S. {Albergo}, D.~{Boyda}, D.~C. {Hackett}, G.~{Kanwar}, K.~{Cranmer},
  S.~{Racani{\`e}re}, D.~{Jimenez Rezende}, and P.~E. {Shanahan},
  ``{Introduction to Normalizing Flows for Lattice Field Theory},'' {\em arXiv
  e-prints}, p.~arXiv:2101.08176, Jan. 2021.

\bibitem{PhysRevB.95.035105}
L.~Huang and L.~Wang, ``Accelerated monte carlo simulations with restricted
  boltzmann machines,'' {\em Phys. Rev. B}, vol.~95, p.~035105, Jan 2017.

\bibitem{2020PhRvE.101e3312M}
B.~{McNaughton}, M.~V. {Milo{\v{s}}evi{\'c}}, A.~{Perali}, and S.~{Pilati},
  ``{Boosting Monte Carlo simulations of spin glasses using autoregressive
  neural networks},'' {\em Phys. Rev. E}, vol.~101, p.~053312, May 2020.

\bibitem{2021arXiv210505650W}
D.~{Wu}, R.~{Rossi}, and G.~{Carleo}, ``{Unbiased Monte Carlo Cluster Updates
  with Autoregressive Neural Networks},'' {\em arXiv e-prints},
  p.~arXiv:2105.05650, May 2021.

\bibitem{PhysRevLett.62.361}
U.~Wolff, ``Collective monte carlo updating for spin systems,'' {\em Phys. Rev.
  Lett.}, vol.~62, pp.~361--364, Jan 1989.

\bibitem{WOLFF1989379}
U.~Wolff, ``Comparison between cluster monte carlo algorithms in the ising
  model,'' {\em Physics Letters B}, vol.~228, no.~3, pp.~379--382, 1989.

\bibitem{PhysRev.185.832}
A.~E. Ferdinand and M.~E. Fisher, ``Bounded and inhomogeneous ising models. i.
  specific-heat anomaly of a finite lattice,'' {\em Phys. Rev.}, vol.~185,
  pp.~832--846, Sep 1969.

\bibitem{Liu}
J.~S. Liu, ``Metropolized independent sampling with comparisons to rejection
  sampling and importance sampling,'' {\em Statistics and Computing}, vol.~6,
  no.~2, pp.~113--119, 1996.

\bibitem{hastings}
W.~K. Hastings, ``{Monte Carlo Sampling Methods Using Markov Chains and Their
  Applications},'' {\em Biometrika}, vol.~57, pp.~97--109, 1970.

\bibitem{10.1214/aos/1176325750}
L.~Tierney, ``{Markov Chains for Exploring Posterior Distributions},'' {\em The
  Annals of Statistics}, vol.~22, no.~4, pp.~1701 -- 1728, 1994.

\bibitem{roberts_rosenthal}
G.~Roberts and J.~Rosenthal, ``{Markov-Chain Monte Carlo: Some Practical
  Implications of Theoretical Results},'' {\em The Canadian Journal of
  Statistics / La Revue Canadienne De Statistique}, vol.~26, pp.~5--20, 1998.

\bibitem{Bialas:2022qbs}
P.~Bia\l{}as, P.~Korcyl, and T.~Stebel, ``Hierarchical autoregressive neural
  networks for statistical systems,'' {\em Comput. Phys. Commun.}, vol.~281,
  p.~108502, 2022.

\bibitem{Sokal1997}
A.~Sokal, ``Monte carlo methods in statistical mechanics: Foundations and new
  algorithms,'' in {\em Functional Integration: Basics and Applications}
  (C.~DeWitt-Morette, P.~Cartier, and A.~Folacci, eds.), (Boston, MA),
  pp.~131--192, Springer US, 1997.

\bibitem{Wolff:2003sm}
U.~Wolff, ``{Monte Carlo errors with less errors},'' {\em Comput. Phys.
  Commun.}, vol.~156, pp.~143--153, 2004.
\newblock [Erratum: Comput.Phys.Commun. 176, 383 (2007)].

\bibitem{Reinforce_book}
A.~B. R.S.~Sutton, {\em Reinforcement Learning: An Introduction.}
\newblock Cambridge, MA: MIT Press, 2018.

\bibitem{minkadivergence}
T.~Minka, ``Divergence measures and message passing,'' {\em technical report,
  Microsoft Research}, 2015.

\bibitem{Hackett:2021idh}
D.~C. Hackett, C.-C. Hsieh, M.~S. Albergo, D.~Boyda, J.-W. Chen, K.-F. Chen,
  K.~Cranmer, G.~Kanwar, and P.~E. Shanahan, ``{Flow-based sampling for
  multimodal distributions in lattice field theory},'' 7 2021.

\end{thebibliography}

% \section{Simulation and implementation details}
% \label{app:simulation}

% \subsection{$\beta$ annealing }

% In order to avoid mode collapse we apply the $\beta$ annealing, namely we start the training process with $\beta_{temp}=0$ (infinite temperature) and then at each epoch the inverse temperature is increased slightly towards aimed $\beta$. We use formula $\beta_{temp}= \beta(1-0.996^{step})$, so that after 1000 steps $\beta_{temp}\approx 0.98\, \beta$.

\end{document}